\documentclass[notoc]{JHEP3}
\usepackage{epsfig}
\usepackage{graphics}
\usepackage{graphicx}
\usepackage{amssymb}
\usepackage{amsmath}
\usepackage{bm}% bold math

\newcommand{\mrad}{{\mathrm{mrad}}}

\newcommand{\sigpolrlhat}{\hat{\sigma}^-}
\newcommand{\sigpolrrhat}{\hat{\sigma}^+}
\newcommand{\spar}{{\stackrel{\rightarrow}{\Rightarrow}}}
\newcommand{\sant}{{\stackrel{\rightarrow}{\Leftarrow}}}

\newcommand{\Sigpolrl}{$\sigma^{\sant}$}
\newcommand{\Sigpolrr}{$\sigma^{\spar}$}

\newcommand{\ra}{\rangle}
\newcommand{\la}{\langle}

\newcommand{\ordera}[1]{$\mathcal{O}(\alpha_s^{#1})$}

\newcommand{\dgg}{$\frac{\Delta g}{g}(x,\mu^2)$}
\newcommand{\dggx}{$\frac{\Delta g}{g}(x)$}
\newcommand{\dggp}{$\la\frac{\Delta g}{g}\ra (p_T)$}
\newcommand{\dggpfact}{\dggp}

\newcommand{\DGG}{$\frac{\Delta g}{g}$}

\def\compass{{\sc Compass}}
\def\hera{{\sc Hera}}
\def\desy{{\sc Desy}}
\def\star{{\sc Star}}
\def\phenix{{\sc Phenix}}
\def\rhic{{\sc Rhic}}
\def\hermes{{\sc Hermes}}
\def\Pythia{{\sc Pythia}}
\def\Jetset{{\sc Jetset}}
\def\msbar{\ensuremath{\overline{\mathrm{MS}}}}
\def\xbj{x_{B}}
\def\2hcut{2.0}
\def\sh2{\hat s^2}
\def\th2{\hat t^2}
\def\uh2{\hat u^2}

\long\def\symbolfootnote[#1]#2{\begingroup%
\def\thefootnote{\fnsymbol{footnote}}\footnote[#1]{#2}\endgroup}

\def\hermesauthor[#1]#2{{#2}$^{\, #1}$}
\def\hermesinstitute[#1]#2{$^{#1\,}$ {#2}\\}

\renewcommand{\thefootnote}{\alph{footnote}}
\def\nowat[#1]#2{\(^,\)\footnote[#1]{#2}}
\title{Leading-Order Determination of the Gluon Polarization
from high-$p_T$ Hadron Electroproduction}

\author{The \hermes\ Collaboration}

\abstract{
Longitudinal double-spin asymmetries of charged hadrons with
high transverse momentum $p_T$ have been measured in electroproduction 
using the \hermes\ detector at \hera.
Processes involving gluons in the nucleon have been enhanced relative to others
by selecting hadrons with $p_T$ typically above 1 GeV.
In this kinematic domain the gluon polarization has been extracted in 
leading order making use of the model embedded in the Monte Carlo 
Generator \Pythia\ 6.2.
The gluon polarization obtained from single inclusive hadrons 
in the $p_T$ range 1 GeV $< p_T <$ 2.5 GeV using a
deuterium target is $\frac{\Delta g}{g}(\langle x\rangle, \langle \mu^2\rangle)=0.049\pm 0.034 (stat)\pm 0.010
(sys\textrm{-}exp)^{+0.126}_{-0.099}(sys\textrm{-}models)$
at a scale $\la\mu^2\ra=1.35~{\rm GeV}^2$ and $\langle x\rangle = 0.22$.
For different final states and kinematic domains, consistent values of \DGG\ 
have been found within statistical uncertainties 
using hydrogen and deuterium targets.}


\preprint{\today}
\begin{document}
%%%%%%%%%%%%%%%%%%%%%%%%%%%%%%%%%%%%%%%%%%%%%%%%%%%%%%%%%%%%%
\section*{The HERMES Collaboration}
{%
\begin{flushleft} 
\bf
\hermesauthor[12,15]{A.~Airapetian},
\hermesauthor[25]{N.~Akopov},
\hermesauthor[5]{Z.~Akopov},
\hermesauthor[6]{E.C.~Aschenauer}\nowat[1]{Now at: Brookhaven National Laboratory, Upton, New York 11772-5000, USA},
\hermesauthor[24]{W.~Augustyniak},
\hermesauthor[25]{R.~Avakian},
\hermesauthor[25]{A.~Avetissian},
\hermesauthor[5]{E.~Avetisyan},
\hermesauthor[17]{S.~Belostotski},
\hermesauthor[10]{N.~Bianchi},
\hermesauthor[16,23]{H.P.~Blok},
\hermesauthor[6]{H.~B\"ottcher},
\hermesauthor[9]{C.~Bonomo},
\hermesauthor[5]{A.~Borissov},
\hermesauthor[18]{V.~Bryzgalov},
\hermesauthor[9]{M.~Capiluppi},
\hermesauthor[10]{G.P.~Capitani},
\hermesauthor[20]{E.~Cisbani},
\hermesauthor[9]{M.~Contalbrigo},
\hermesauthor[9]{P.F.~Dalpiaz},
\hermesauthor[5,15]{W.~Deconinck}\nowat[2]{Now at:Massachusetts Institute of Technology, Cambridge, Massachusetts 02139, USA},
\hermesauthor[2]{R.~De~Leo},
\hermesauthor[16]{M.~Demey}
%\nowat[3]{Now at: ??? to be added},
\hermesauthor[15,5]{L.~De~Nardo},
\hermesauthor[10]{E.~De~Sanctis},
\hermesauthor[14,8]{M.~Diefenthaler},
\hermesauthor[10]{P.~Di~Nezza},
\hermesauthor[16]{J.~Dreschler}
%\nowat[4]{Now at: ???, to be added},
\hermesauthor[12]{M.~D\"uren},
\hermesauthor[12]{M.~Ehrenfried}
%\nowat[5]{Now at: Siemens AG Molecular Imaging, 91052 Erlangen, Germany},
\hermesauthor[25]{G.~Elbakian},
\hermesauthor[4]{F.~Ellinghaus}\nowat[3]{Now at:Institut f\"ur Physik, Universit\"at Mainz, 55128 Mainz, Germany},
\hermesauthor[11]{U.~Elschenbroich}
%\nowat[7]{Now at: Siemens AG, 91052 Erlangen, Germany},
\hermesauthor[6]{R.~Fabbri},
\hermesauthor[10]{A.~Fantoni},
\hermesauthor[21]{L.~Felawka},
\hermesauthor[20]{S.~Frullani},
\hermesauthor[11,6]{D.~Gabbert},
\hermesauthor[18]{G.~Gapienko},
\hermesauthor[18]{V.~Gapienko},
\hermesauthor[20]{F.~Garibaldi},
\hermesauthor[5,17,21]{G.~Gavrilov},
\hermesauthor[25]{V.~Gharibyan},
\hermesauthor[5,9]{F.~Giordano},
\hermesauthor[15]{S.~Gliske},
\hermesauthor[6]{H.~Guler}
%\nowat[8]{Now at: Laboratoire Leprince-Ringuet, CNRS/IN2P3, Ecole Polytechnique, 91128 Palaiseau, France},
\hermesauthor[10]{C.~Hadjidakis}\nowat[4]{Now at: IPN (UMR 8608) CNRS/IN2P3 - Universit\'e Paris-Sud, 91406 Orsay, France},
\hermesauthor[5]{M.~Hartig}\nowat[5]{Now at: Institut f\"ur Kernphysik, Universit\"at Frankfurt a.M., 60438 Frankfurt a.M., Germany},
\hermesauthor[10]{D.~Hasch},
\hermesauthor[22]{T.~Hasegawa},
\hermesauthor[13]{G.~Hill},
\hermesauthor[6]{A.~Hillenbrand},
\hermesauthor[13]{M.~Hoek},
\hermesauthor[5]{Y.~Holler},
\hermesauthor[11]{B.~Hommez},
\hermesauthor[6]{I.~Hristova},
\hermesauthor[18]{A.~Ivanilov},
\hermesauthor[1]{H.E.~Jackson},
\hermesauthor[13]{R.~Kaiser},
\hermesauthor[13,12]{T.~Keri},
\hermesauthor[4]{E.~Kinney},
\hermesauthor[17]{A.~Kisselev},
\hermesauthor[6]{M.~Kopytin},
\hermesauthor[18]{V.~Korotkov},
\hermesauthor[17]{P.~Kravchenko},
\hermesauthor[2]{L.~Lagamba},
\hermesauthor[14]{R.~Lamb},
\hermesauthor[16]{L.~Lapik\'as},
\hermesauthor[13]{I.~Lehmann},
\hermesauthor[9]{P.~Lenisa},
\hermesauthor[6]{P.~Liebing}\nowat[6]{Now at: Institute of Environmental Physics and Remote Sensing, University of Bremen, 28334 Bremen, Germany},
\hermesauthor[14]{L.A.~Linden-Levy},
\hermesauthor[15]{W.~Lorenzon},
\hermesauthor[22]{X.-R.~Lu}
%\nowat[12]{Now at: Graduate University of Chinese Academy of Sciences, Beijing 100049, China} ,
\hermesauthor[11]{B.~Maiheu},
\hermesauthor[14]{N.C.R.~Makins},
\hermesauthor[24]{B.~Marianski},
\hermesauthor[25]{H.~Marukyan},
\hermesauthor[16]{V.~Mexner},
\hermesauthor[21]{C.A.~Miller},
\hermesauthor[22]{Y.~Miyachi},
\hermesauthor[10]{V.~Muccifora},
\hermesauthor[13]{M.~Murray},
\hermesauthor[5,8]{A.~Mussgiller},
\hermesauthor[2]{E.~Nappi},
\hermesauthor[17]{Y.~Naryshkin},
\hermesauthor[8]{A.~Nass},
\hermesauthor[6]{M.~Negodaev},
\hermesauthor[6]{W.-D.~Nowak},
\hermesauthor[9]{L.L.~Pappalardo},
\hermesauthor[12]{R.~Perez-Benito},
\hermesauthor[8]{N.~Pickert}
%\nowat[13]{Now at:Siemens AG, 91301 Forchheim, Germany},
\hermesauthor[8]{M.~Raithel},
\hermesauthor[8]{D.~Reggiani}
\hermesauthor[1]{P.E.~Reimer},
\hermesauthor[16]{A.~Reischl}
\hermesauthor[10]{A.R.~Reolon},
\hermesauthor[6]{C.~Riedl},
\hermesauthor[8]{K.~Rith},
\hermesauthor[5]{S.E.~Rock}\nowat[7]{Present affiliation: SLAC National Accelerator Laboratory, Menlo Park, California 94025, USA},  
\hermesauthor[13]{G.~Rosner},
\hermesauthor[5]{A.~Rostomyan},
\hermesauthor[1,14]{J.~Rubin},
\hermesauthor[18]{Y.~Salomatin},
\hermesauthor[19]{A.~Sch\"afer},
\hermesauthor[6,22]{G.~Schnell},
\hermesauthor[5]{K.P.~Sch\"uler},
\hermesauthor[13]{B.~Seitz},
\hermesauthor[13]{C.~Shearer}
\hermesauthor[22]{T.-A.~Shibata},
\hermesauthor[7]{V.~Shutov},
\hermesauthor[9]{M.~Stancari},
\hermesauthor[9]{M.~Statera},
\hermesauthor[16]{J.J.M.~Steijger},
\hermesauthor[6]{J.~Stewart}
%\nowat[1]{Now at: Brookhaven National Laboratory, Upton, New York 11772-5000, USA},
\hermesauthor[8]{F.~Stinzing},
\hermesauthor[25]{S.~Taroian},
\hermesauthor[18]{B.~Tchuiko},
\hermesauthor[24]{A.~Trzcinski},
\hermesauthor[11]{M.~Tytgat},
\hermesauthor[11]{A.~Vandenbroucke}
%\nowat[18]{Now at: Dept of Radiology, Stanford University, School of Medicine, Stanford, California 94305-5105, USA},
\hermesauthor[16]{P.B.~van~der~Nat}
\hermesauthor[16]{G.~van~der~Steenhoven}
%\nowat[20]{Now at: Faculty of Science and Technology, University Twente, 7500 AE Enschede, The Netherlands},
\hermesauthor[11]{Y.~Van~Haarlem}\nowat[8]{Now at: Carnegie Mellon University, Pittsburgh, Pennsylvania 15213, USA},
\hermesauthor[11]{C.~Van~Hulse},
\hermesauthor[5]{M.~Varanda}
%\nowat[22]{Now at: Faculty of Science, Universidade de Lisboa, 1749-016 Lisboa, Portugal},
\hermesauthor[17]{D.~Veretennikov},
\hermesauthor[2]{I.~Vilardi}
%\nowat[23]{Now at: IRCCS Multimedica Holding S.p.A., 20099 Sesto San Giovanni (MI), Italy},
\hermesauthor[8]{C.~Vogel}
%\nowat[24]{Now at:AREVA NP GmbH, 91058 Erlangen, Germany},
\hermesauthor[3]{S.~Wang},
\hermesauthor[6,8]{S.~Yaschenko},
\hermesauthor[3]{H.~Ye},
\hermesauthor[5]{Z.~Ye}
%\nowat[25]{Now at: Fermi National Accelerator Laboratory, Batavia, Illinois 60510, USA},
\hermesauthor[12]{W.~Yu},
\hermesauthor[8]{D.~Zeiler},
\hermesauthor[11]{B.~Zihlmann}
%\nowat[26]{Now at: Thomas Jefferson National Accelerator Facility, Newport News, Virginia 23606, USA},
\hermesauthor[24]{P.~Zupranski}
\end{flushleft} 
%\end{center}
}
%-- HERMES Institutes
\bigskip
{\it
%\begin{center}
\begin{flushleft} 
\hermesinstitute[1]{Physics Division, Argonne National Laboratory, Argonne, Illinois 60439-4843, USA}
\hermesinstitute[2]{Istituto Nazionale di Fisica Nucleare, Sezione di Bari, 70124 Bari, Italy}
\hermesinstitute[3]{School of Physics, Peking University, Beijing 100871, China}
\hermesinstitute[4]{Nuclear Physics Laboratory, University of Colorado, Boulder, Colorado 80309-0390, USA}
\hermesinstitute[5]{DESY, 22603 Hamburg, Germany}
\hermesinstitute[6]{DESY, 15738 Zeuthen, Germany}
\hermesinstitute[7]{Joint Institute for Nuclear Research, 141980 Dubna, Russia}
\hermesinstitute[8]{Physikalisches Institut, Universit\"at Erlangen-N\"urnberg, 91058 Erlangen, Germany}
\hermesinstitute[9]{Istituto Nazionale di Fisica Nucleare, Sezione di Ferrara and Dipartimento di Fisica, Universit\`a di Ferrara, 44100 Ferrara, Italy}
\hermesinstitute[10]{Istituto Nazionale di Fisica Nucleare, Laboratori Nazionali di Frascati, 00044 Frascati, Italy}
\hermesinstitute[11]{Department of Subatomic and Radiation Physics, University of Gent, 9000 Gent, Belgium}
\hermesinstitute[12]{Physikalisches Institut, Universit\"at Gie{\ss}en, 35392 Gie{\ss}en, Germany}
\hermesinstitute[13]{Department of Physics and Astronomy, University of Glasgow, Glasgow G12 8QQ, United Kingdom}
\hermesinstitute[14]{Department of Physics, University of Illinois, Urbana, Illinois 61801-3080, USA}
\hermesinstitute[15]{Randall Laboratory of Physics, University of Michigan, Ann Arbor, Michigan 48109-1040, USA }
\hermesinstitute[16]{National Institute for Subatomic Physics (Nikhef), 1009 DB Amsterdam, The Netherlands}
\hermesinstitute[17]{Petersburg Nuclear Physics Institute, Gatchina, Leningrad region 188300, Russia}
\hermesinstitute[18]{Institute for High Energy Physics, Protvino, Moscow region 142281, Russia}
\hermesinstitute[19]{Institut f\"ur Theoretische Physik, Universit\"at Regensburg, 93040 Regensburg, Germany}
\hermesinstitute[20]{Istituto Nazionale di Fisica Nucleare, Sezione Roma 1, Gruppo Sanit\`a and Physics Laboratory, Istituto Superiore di Sanit\`a, 00161 Roma, Italy}
\hermesinstitute[21]{TRIUMF, Vancouver, British Columbia V6T 2A3, Canada}
\hermesinstitute[22]{Department of Physics, Tokyo Institute of Technology, Tokyo 152, Japan}
\hermesinstitute[23]{Department of Physics, VU University, 1081 HV Amsterdam, The Netherlands}
\hermesinstitute[24]{Andrzej Soltan Institute for Nuclear Studies, 00-689 Warsaw, Poland}
\hermesinstitute[25]{Yerevan Physics Institute, 375036 Yerevan, Armenia}
\end{flushleft} 
%\end{center}
}

\newpage
%%%%%%%%%%%%%%%%%%%%%%%%%%%%%%%%%%%%%%%%%%%%%%%%%%%%%%%%%%%%%

%%%%%%%%%%%%%%%%%%%%%%%%%%%%%%%%%%%%%%%%%%%%%%%%%%%%%%%%%%%%%
%%%%%%%%%%%%%%%%%%%%%  Introduction   %%%%%%%%%%%%%%%%%%%%%%%
%%%%%%%%%%%%%%%%%%%%%%%%%%%%%%%%%%%%%%%%%%%%%%%%%%%%%%%%%%%%%
\section{Introduction}
\label{into}
%%%%%%%%%%%%%%%%%%%%%%%%%%%%%%%%%%%%%%%%%%%%%%%%%%%%%%%%%%%%%
In recent years a major goal in the study of Quantum Chromo-Dynamics (QCD)  
has been the detailed investigation of the spin structure of the nucleon
and the determination of the partonic composition of its spin
projection~\cite{IFM:spin1990}
\begin{equation}
\label{eq:spintot}
 \frac{1}{2} = S_z = \frac{1}{2}\cdot \Delta \Sigma + \Delta G + L_z^q + L_z^G.
\end{equation}
Here $\Delta \Sigma$ is the contribution of the quark
and anti-quark helicities, $\Delta G$ is the contribution of the
gluon helicity, and $L_z^q$ and $L_z^G$ are the quark and gluon orbital angular
momenta, respectively, in a reference system where the nucleon has very large 
longitudinal momentum. The individual terms in the sum depend on the scale 
$\mu^2$ and the renormalization scheme.
Recent results from experiments~\cite{hermes:g1-06,compass:2006g1}
and fits in next-to-leading order (NLO)
QCD~\cite{Blumlein:2002be,AAC:2006,th:DSSV2008,pdf:lss2005} 
to \mbox{helicity-dependent} inclusive Deep-Inelastic Scattering (DIS) 
data~\cite{hermes:g1-06,compass:2006g1,hermes:air2005,emc:g1,e142:g1n_g2n,%
hermes:g1n,e154:g1n,smc:g1p-g1d,e143:long,Anthony:1999rm,%
e155:g1p_g1d,compass:g1,g1:JLAB-A04}
yield a value of $\Delta \Sigma \sim 0.2 - 0.4$ at 
$\mu^2=4$ GeV$^2$ in the \msbar\ scheme~\cite{th:bardeen}.
In contrast to the quark helicity distributions, the knowledge of the 
gluon helicity distribution function is still limited. 
There are no direct experimental determinations of parton orbital angular 
momenta. 
Most of the existing knowledge about $\Delta G(\mu^2)$ originates 
from next-to-leading order perturbative QCD (pQCD) fits to the 
\mbox{helicity-dependent} structure function $g_1(\xbj,Q^2)$ of the 
nucleon, where $\xbj$ is the Bjorken scaling variable, which is in 
leading-order (LO) identified with the longitudinal parton momentum fraction $x$ 
in the nucleon. In DIS the renormalization and factorization scales 
$\mu^2$ are set equal to the photon virtuality $Q^2$. Because
the virtual photon does not couple directly to gluons 
(see Fig.~\ref{fg:qcddiag}b), $g_1(\xbj,Q^2)$ is only weakly 
sensitive to gluons through the DGLAP evolution 
\cite{dglap:do,dglap:gl,dglap:ap} of the helicity-dependent Parton Distribution 
Functions (PDFs).
At next-to-leading order pQCD, additional sensitivity to gluons 
arises from the Photon-Gluon Fusion (PGF) subprocess 
(see Fig.~\ref{fg:qcddiag}b). 
However, the limitations on the precision and kinematic range in $\xbj$
and $Q^2$ of the $g_1$ measurements result in large experimental and 
theoretical uncertainties on the determination of the gluon helicity 
distribution function $\Delta g(x,\mu^2)$. 
Results for $\Delta G(\mu^2) = \int_0^1 \Delta g(x,\mu^2) dx$ from recent 
pQCD fits to inclusive DIS data 
\cite{Blumlein:2002be,AAC:2006,th:DSSV2008,pdf:lss2005}
are typically of order 0.5 with uncertainties up to $\pm 1$.

An alternative constraint on the extraction of $\Delta G(\mu^2)$ in NLO 
pQCD fits comes 
from the measurements of double-spin asymmetries in production of inclusive 
$\pi^0$ mesons or jets  with high transverse
momentum in polarized proton-(anti-)proton scattering. 
First measurements were performed by E704~\cite{exp:e704} and more recent
data were obtained by \phenix\ \cite{phenix:Adl} and \star\ \cite{star2006} 
at \rhic. 
The inclusion of the RHIC-data in recent NLO pQCD 
fits~\cite{th:DSSV2008} improves the accuracy on $\Delta g$ significantly. 
One finds $|\Delta g(x,Q^2)|$ smaller than 0.1, with a possible 
node in the distribution. This is driven mainly by the RHIC
data, which constrain the magnitude of $\Delta g(x)$ for 
$0.05 < x < 0.2$, but cannot determine its sign as they
mainly probe the product of the gluon helicity distribution at two $x$ values.

%*************************************************************************
\FIGURE{
 \includegraphics*[width=13.5cm]{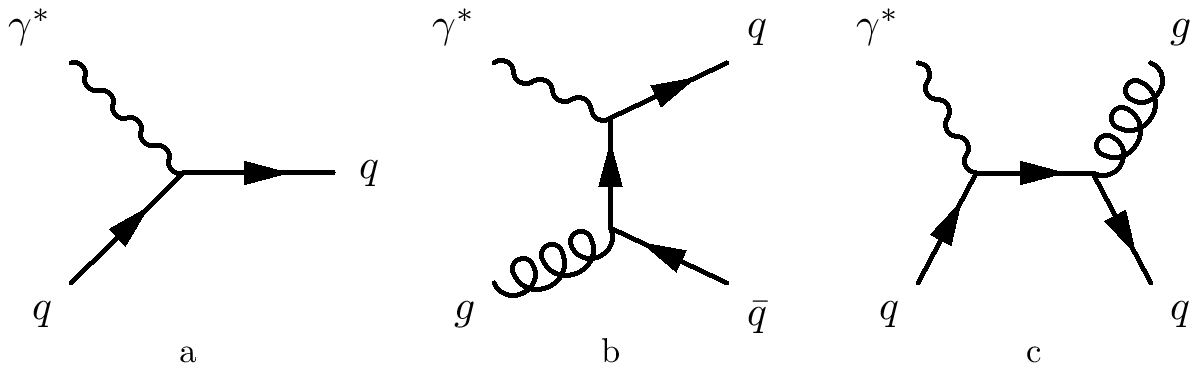}
 \caption{\label{fg:qcddiag}Feynman diagrams for hard subprocesses: 
  a) \ordera{0} DIS, b) \ordera{1} Photon-Gluon Fusion, and 
  c) \ordera{1} QCD Compton scattering.}
}
%*************************************************************************

In order to increase the sensitivity to $\Delta g(x,\mu^2)$ 
in lepton-nucleon scattering, other observables besides the inclusive 
helicity-dependent structure function have been studied.
These observables are expected to include a direct contribution from gluons.
For example, in hadron leptoproduction this gluonic contribution can be 
relatively enhanced by 
detecting charmed hadrons, or inclusive hadrons or hadron pairs
at high transverse momenta $p_T$.

Charmed hadron electroproduction is a suitable channel because it is 
dominated by  the PGF subprocess~\cite{compass:dg2008} and a hard 
scale is introduced by the mass of the charm-quark pair, which makes
pQCD calculations of this process possible.  
For light final state quarks, the selection of hadrons 
with high $p_T$ enhances the relative contribution of the gluon
subprocesses and the relevant transverse momenta provide the scale
(see Sect.~\ref{sec:MCevents}). 
In the high $p_T$ domain other calculable hard pQCD subprocesses such as 
QCD Compton (QCDC) scattering (see Fig.~\ref{fg:qcddiag}c) are 
relatively enhanced as well, whereas soft, non-perturbative processes are 
suppressed. Charm electroproduction is being investigated by 
\compass~\cite{compass:dg2008,compass:proposal}. 
Inclusive single hadron leptoproduction was studied by 
E155~\cite{exp:e155}.
Hadron-pair leptoproduction at high $p_T$ was studied by 
\hermes~\cite{hermes:old-glue}, SMC~\cite{smc:glue} and
\compass~\cite{compass:dg2006,compass:DIS08}. 
For these experiments high $p_T$ is in the range from one to a few GeV.

Throughout this paper, the term ``LO'' is applied to all leading order
subprocesses contributing to hadron production at nonzero $p_T$. These are
the tree level processes at $\mathcal{O}(\alpha_s^1)$, but also the 
quark scattering process $\gamma^*q\rightarrow q$ (DIS) at 
$\mathcal{O}(\alpha_s^0)$. While the former processes involve hard gluons, and
can therefore involve substantial parton transverse momentum 
${\hat p}_T$ in the hard scattering, in the latter process 
${\hat p}_T$ is equal to zero, but hadrons acquire $p_T$ from soft initial 
and final state 
radiation. This paper presents the LO extraction of the gluon polarization 
\dggx\ from longitudinal \mbox{double-spin} asymmetries of charged inclusive 
hadrons
measured in electroproduction using a deuterium target by \hermes\ at \hera. 
The contributions of signal and background have been
determined by a \Pythia\ Monte Carlo simulation, which 
includes LO pQCD as well as non-perturbative 
subprocesses. Consistency checks have been performed for different kinematic 
regions, different final states and using data from a hydrogen target. 
The data taken with the deuterium target correspond to an
integrated luminosity three times larger than that taken with the hydrogen
target, see table~\ref{tb:tar}. Compared to the previous 
\hermes\ publication~\cite{hermes:old-glue}, which used measurements
of hadron pairs of opposite charge from a hydrogen target, this analysis includes 
a much larger sample of  single hadrons, and a significantly more comprehensive 
treatment of the underlying physics 
processes~\cite{hermes:liebing,hermes:mexner}.

This paper is organized as follows:
In Sect.~\ref{sec:exp} the experimental method is described, in 
Sect.~\ref{sec:ExpResults} the asymmetry results are given, in
Sect.~\ref{sec:physics} the determination of
\DGG\ with a description of the physics model of the reactions is discussed,
in Sect.~\ref{sec:pythia} the \Pythia\ Monte Carlo simulation is described,
in Sect.~\ref{sec:DG} the determination of \dgg\ is explained, and in
Sect.~\ref{sec:conclusion} the summary and conclusions are given.

%%%%%%%%%%%%%%%%%%%%%%%%%%%%%%%%%%%%%%%%%%%%%%%%%%%%%%
%%              The HERMES Experiment               %%
%%%%%%%%%%%%%%%%%%%%%%%%%%%%%%%%%%%%%%%%%%%%%%%%%%%%%%
%%%%%%%%%%%%%%%%%%%%%%%%%%%%%%%%%%%%%%%%%%%%%%%%%%%%%%%%%%%%%%%%%%%%%%%%%%%%%%
\section{The \hermes\ Experiment}
\label{sec:exp}
%%%%%%%%%%%%%%%%%%%%%%%%%%%%%%%%%%%%%%%%%%%%%%%%%%%%%%%%%%%%%%%%%%%%%%%%%%%%%%
Positrons of momentum 27.6 GeV were stored in the \hera\ lepton ring 
at {\sc Desy}. The initially unpolarized beam was transversely polarised by 
an asymmetry in the emission of synchrotron radiation associated with a spin 
flip (Sokolov-Ternov mechanism \cite{ST:mech}). The polarization
was rotated to the longitudinal direction for passage through an gaseous
internal fixed target of longitudinally nuclear-polarized atoms. The scattered positron 
and  hadrons produced were detected in a forward magnetic spectrometer.
The beam helicity was reversed periodically. The beam polarization was measured 
continuously by two independent polarimeters using Compton backscattering of 
circularly polarized laser light~\cite{hermes:lpol,hermes:tpol2}. 
The average beam polarization for the data used in this analysis is shown 
in Tab.~\ref{tb:tar}.
The target~\cite{hermes:Target} consisted of longitudinally nuclear-polarized 
pure atomic hydrogen or deuterium gas in an open-ended 40\,cm long  
storage cell. The cell was fed by an atomic-beam source based on Stern-Gerlach
separation combined with radio-frequency transitions of atomic hyperfine 
states~\cite{hermes:ABS}.
The sign of the nuclear polarization of the atoms was chosen randomly 
every 60 s (90 s) for the hydrogen (deuterium) target.
The polarization and the atomic fraction inside the target cell were
continuously measured~\cite{hermes:BRP,hermes:TGA}.  
The average values of the target polarization for both hydrogen and 
deuterium data are shown in  Tab.~\ref{tb:tar}.
The luminosity was measured by detecting 
$e^+e^-$ pairs originating from Bhabha scattering of the beam positrons off
electrons in the target atoms, and also $\gamma\gamma$ pairs from $e^+e^-$ 
annihilations~\cite{Benisch:2001rr}. 

%*****************************************************************************
\TABLE{
  \caption{\label{tb:tar}Integrated luminosities, average beam and target 
           polarizations for the data used in this analysis.}
  \begin{tabular}{|l|c|c|c|c|}

\hline
 Year     & Target & Luminosity &Average Beam     &Average Target   \\ 
          &        & pb$^{-1}$   & Polarization    & Polarization \\ 
\hline
  1996 \rule{0ex}{3ex}&  H & 12.6 & $0.528\pm0.018 $ & $ 0.759\pm0.032$\\
  1997                &  H & 37.3 & $0.531\pm0.018 $ & $ 0.851\pm0.032$\\
  2000                &  D &138.7 & $0.533\pm0.010 $ & $ 0.846\pm0.030$\\ 
\hline
  \end{tabular}
}
%*****************************************************************************

The \hermes\ spectrometer~\cite{hermes:spectr} consisted of two identical 
halves separated by a horizontal flux diversion plate,  
which limited the minimum detected angle. The geometrical acceptance was
$\pm 170~\mrad$ in the horizontal (bending) plane and between
$\pm (40 - 140) ~\mrad$ in the vertical plane resulting in a
range of polar angles between $40 ~\mrad$ and $220 ~\mrad$. Each half was
instrumented with 3 planes of hodoscopes, 36 planes of drift chambers, 
and 9 planes of proportional chambers. The particle identification system
consisted of an electromagnetic calorimeter, a pre-shower hodoscope,
a transition-radiation detector, and a {\v C}erenkov detector.
Detailed descriptions of these components can be found in Refs.~
\cite{hermes:spectr,hermes:FCs,hermes:MCs,hermes:RICH,hermes:BCs,hermes:Calo}.
Positrons within the acceptance could be separated from hadrons
with an efficiency exceeding 98\% and a hadron contamination of less than 1\%.  

The main \hermes\ physics trigger was formed by a coincidence of hits in
the hodoscopes in the front and back regions of the spectrometer 
with the requirement of an energy deposit above 1.4 GeV in the calorimeter.
This trigger was almost 100\% efficient for positrons with energies above 
threshold. Events with no positron in the acceptance were recorded
using a mixture of the main trigger and another trigger formed
by a coincidence between the hodoscopes and two tracking planes, requiring 
that there is at least one charged track.
The influence of trigger efficiencies on the analysis has been studied in
\cite{hermes:liebing}.

%%%%%%%%%%%%%%%%%%%%%%%%%%%%%%%%%%%%%%%%%%%%%%%%%%%%%%%%
%%%%%%%%%%%%%%%%%  Experimental Results %%%%%%%%%%%%%%%%%%
%%%%%%%%%%%%%%%%%%%%%%%%%%%%%%%%%%%%%%%%%%%%%%%%%%%%%%%%
\section{Experimental Results}
\label{sec:ExpResults}
%%%%%%%%%%%%%%%%%%%%%%%%%%%%%%%%%%%%%%%%%%%%%%%%%%%%%%%%
The ratio \dgg\ of helicity-dependent to helicity-averaged gluon distributions,
\textit{i.e.} the gluon polarization, is determined by measuring the 
double-spin cross section asymmetry of one or two high-$p_T$ inclusive 
hadrons produced in the scattering of longitudinally polarized positrons 
incident on the longitudinally polarized target.  
The definitions of the kinematic variables in
electroproduction used in this paper are shown in Tab.~\ref{tb:kin}.
The longitudinal double-spin cross section asymmetry is given by the ratio of
helicity-dependent to helicity-averaged cross sections
$A $=$\Delta\sigma/(2\sigma)$, where
$\sigma$ = (\Sigpolrl $+$ \Sigpolrr)/2, $\Delta\sigma =$\Sigpolrl $-$ \Sigpolrr,
and the single (double) arrows denote the relative alignment of longitudinal
spins of the lepton (nucleon) with respect to the lepton beam direction.

The data for this analysis were collected in 1996, 1997, and 2000 
(see Tab.~\ref{tb:tar}). 
The analysis presented in this paper includes all (unidentified)
charged hadrons. Separate asymmetries are given for each charge,
target, and event category. 
%*************************************************************************
\TABLE{
 \caption{\label{tb:kin} Definition of kinematic variables.} 
  \begin{tabular}{ll} \hline
   $k=(E,\vec{k})$, $k'=(E',\vec{k'})$ & 4--momenta of the initial 
                                          and final state leptons \\
   $\theta,\; \phi$ & Polar and azimuthal angle of the scattered positron\\
   $M$      & Mass of the initial target nucleon\\
   $q=(E-E',\vec{k}-\vec{k'})$ & 4--momentum of the virtual photon \\
   $Q^2 = -q^2 \stackrel{\mathrm{lab}}{\approx} 4EE'\sin^2\frac{\theta}{2}$ &  
          Negative squared 4-momentum transfer\\
   ${\nu=\frac{P\cdot q}{M}\stackrel{\mathrm{lab}}{=} E-E'}$ & Energy of the virtual photon \\
   $x=\frac{Q^2}{2\,P\cdot q} \stackrel{\mathrm{lab}}{=} \frac{Q^2}{2\,M\nu} $ & Bjorken scaling variable\\
   $y=\frac{P\cdot q}{P\cdot k} \stackrel{\mathrm{lab}}{=} \frac{\nu}{E}$ & Fractional energy of the virtual photon \\
   $W^2=(P+q)^2$\\\hspace*{2em} $ \stackrel{\mathrm{lab}}{=}M^2+2M\nu-Q^2$ & Squared invariant mass of the virtual-photon nucleon system \\
   $ p=(E_h,\vec{p})$ &  4-momentum of a hadron in the final state \\
   $ p_T $ &  Transverse momentum of a hadron \\
   $ p_{T(\gamma^*)}$   &  $p_T$  with respect to the virtual photon \\
   $ p_{T(beam)}$  &  $p_T$  with respect to the incoming positron \\
   $p_T^{frag}$ & Transverse hadron  momentum from fragmentation\\
   $\sum{p_{T(beam)}^2}$ &   For two hadrons:~$(p_{T(beam)}^{h1})^2 + (p_{T(beam)}^{h2})^2$ \\
   $ z = \frac{P\cdot P_h }{P\cdot q} \stackrel{\mathrm{lab}}{=} \frac{E_h}{\nu}$ & Fractional energy of the final state hadron\\
   $x$           & Parton momentum fraction \\
   $\hat s=(p_a+p_b)^2$ & Mandelstam variable for partonic process $ab\rightarrow cd$\\
   $\hat t=(p_a-p_c)^2$ & Mandelstam variable for partonic process $ab\rightarrow cd$\\
   $\hat u=(p_b-p_c)^2$ & Mandelstam variable for partonic process $ab\rightarrow cd$\\
   $\mu^2$       & pQCD scale\\
%   $\hat{Q}^2$   & Scale of the partonic hard subprocess \\
   $\hat{p}_T\ (=\sqrt{\frac{\hat{u}\hat{t}}{\hat{s}}}$~ for $m=0$) & 
                Transverse momentum of final state partons \\ [-1.4ex]
               & in the CM-system of the hard subprocess \\
   $k_T$ & Intrinsic transverse momentum of partons \\ [-1.4ex]
         & in the nucleon and photon \\
                 
 \hline
\end{tabular}
}
%*************************************************************************

%%%%%%%%%%%%%%%%%%%%%%%%%%%%%%%%%%%%%%%%%%%%%%%%%%%%%%%%%%%%%%%%%%%%%%%%%%%%
\subsection{Event categories}
\label{sec:kinematics}
%%%%%%%%%%%%%%%%%%%%%%%%%%%%%%%%%%%%%%%%%%%%%%%%%%%%%%%%%%%%%%%%%%%%%%%%%%%%

Simulations indicate that subprocesses involving hard gluons are relatively
enhanced by measuring hadrons with high $p_T$ with respect to the virtual 
photon direction ($p_{T(\gamma^*)}$). 
Correlations between hadrons in an event may also enhance the signal. 
Events are categorized by the number of hadrons observed in an event and
whether kinematic information on the scattered positron is available or not.
Each possible combination of two hadrons is counted as a separate event in the 
pairs category. The categories are defined in detail as follows:

\begin{itemize}
\item{\bf `anti-tagged' single inclusive hadrons:}
Events with leptons in the acceptance were not included in this category.
The hadron transverse momentum $p_{T(beam)}$ was measured with respect to the
beam direction as the direction of the virtual photon is unknown.  
In most cases, the undetected positron had a small scattering angle 
(and hence $Q^2$ is small) and stayed inside the beam pipe.
The difference between $p_{T(beam)}$ and $p_{T(\gamma^*)}$ is then very small.
However, the positron could also escape the detector acceptance because of a 
large scattering angle, in which case $Q^2$ was large.
The large angle of the virtual photon with respect to the beam axis
results in a significantly larger $p_{T(beam)}$ than 
$p_{T(\gamma^*)}$ of the hadron.  
Although these events with large $Q^2$ are rare, they can account for a 
significant fraction of the hadrons at high $p_{T(beam)}$.  For $p_T >1.0$~GeV 
the deuteron (proton) data sample in this category contains 1272k (419k) hadrons.
\item {\bf `tagged' single inclusive hadrons:}
The scattered positron has been detected with $Q^2>0.1~{\rm GeV}^2$, 
$W^2>4{\rm ~GeV}^2$, and $y<0.95$. 
The hadron transverse momentum $p_{T(\gamma^*)}$ is measured with 
respect to the virtual photon direction.
For $p_T >1$~GeV this deuteron (proton) data sample contains 53k (19k) hadrons.
\item {\bf inclusive pairs of hadrons:}
The hadron pair sample consists of all pairs of charged hadrons with 
$p_{T(beam)}>0.5$ GeV. The transverse momentum $p_{T(beam)}$
is measured with respect to the beam direction, because only in 10\% of 
the events the positron was detected. With the  additional requirement 
$\sum{p_{T(beam)}^2} >\2hcut$~GeV$^2$ 
the deuteron (proton) data sample contains 60k (20k) hadron pairs. 
With this requirement applied, 6\% of the anti-tagged inclusive hadrons with 
$p_{T(beam)}>1.0$ GeV are contained within the pairs sample.
\end{itemize}
For all three  categories, \hermes\ data are available for various combinations
of target and/or hadron charge detected. 
As the samples differ in the hard subprocess and final state kinematics 
and fractions of contributing subprocesses, the corresponding results for the gluon 
polarization \dgg\ provide a measure of the consistency of the extraction.
The final result for  \dgg\ is obtained from the 
anti-tagged inclusive hadrons originating from a deuterium target. 
The other data samples have too small a statistical power to justify
carrying out the extensive analysis needed to obtain the systematic
uncertainties. 

%%%%%%%%%%%%%%%%%%%%%%%%%%%%%%%%%%%%%%%%%%%%%%%%%%%%%%%%%%%%%%%%%%%%%%%%%%%%%%
\subsection{Asymmetry results}
\label{sec:asym_res}
%%%%%%%%%%%%%%%%%%%%%%%%%%%%%%%%%%%%%%%%%%%%%%%%%%%%%%%%%%%%%%%%%%%%%%%%%%%%%%
The double-spin asymmetry  measured is given by
%*****************************************************************************
\begin{equation}
  A_{meas} \equiv A_{||}=\frac{N^\sant L^\spar - N^\spar L^\sant}
  {N^\sant L_P^\spar + N^\spar L_P^\sant}\, . 
  \label{eq:aparallel}
\end{equation}
%*****************************************************************************
Here $N^\spar$ ($N^\sant$) is the  number of hadrons or hadron pairs 
for target spin orientation parallel (anti-parallel) to the beam spin 
orientation, $L^\spar$ ($L^\sant$) is the corresponding integrated luminosity, 
and $L_P^\spar$ ($L_P^\sant$) is the integrated luminosity weighted by the live-time fraction and the absolute values of beam and target polarizations. 
There is a small background ($<0.1\%$) from positrons misidentified as
hadrons (and vice versa).  In the tagged category  a correction was applied for
an approximately $5\%$ contribution of positrons  
originating from  charge-symmetric processes. 

The asymmetries for the anti-tagged and tagged categories are shown as 
a function of transverse momentum in Figs.~\ref{fg:1h_at_asym_dp} and 
\ref{fg:1h_t_asym_dp}, respectively, and listed in 
tables~\ref{tb:Anti-tagged} - \ref{tb:Tagged}.
The asymmetry of the pairs is presented as a function of the minimum
requirement,  $(\sum{p_{T(beam)}^2})_{min}$, in 
Fig.~\ref{fg:2h_asym_dp} and in table~\ref{tb:2hasym}.
The considerably different values of the asymmetries in the different 
categories, charges and targets are due to the different underlying mixtures of 
subprocesses and of quark content, as discussed in Sect.~\ref{sec:MCevents}. 
%*************************************************************************
\FIGURE[h]{
  \includegraphics*[width=13.5cm]{./figs/asym_photo_paper_hrc.epsi}
  \caption{\label{fg:1h_at_asym_dp} 
  Measured asymmetry for the anti-tagged category of events for positive 
  (left) and negative (right) inclusive hadrons from hydrogen (top) and 
  deuterium (bottom) targets as a function of $p_{T(beam)}.$ 
  The uncertainties are statistical only. There is an overall normalization 
  uncertainty of 5.2\% (3.9\%) for hydrogen (deuterium).
  The curves show the Monte Carlo asymmetries for three different fixed values 
  assumed for the gluon polarization.}
}
%*************************************************************************
\FIGURE[h]{
  \includegraphics*[width=13.5cm]{./figs/asym_semi_paper_hrc.epsi}
  \caption{\label{fg:1h_t_asym_dp} 
  Measured asymmetry for the tagged category of events for positive (left) and 
  negative (right) inclusive hadrons from hydrogen (top) and 
  deuterium (bottom) targets as a function of $p_{T(\gamma^*)}.$ 
  The uncertainties are statistical only. There is an
  overall experimental normalization  uncertainty of 5.2\% (3.9\%) 
  for hydrogen (deuterium). The curves show the Monte Carlo asymmetries for 
  three different fixed values assumed for the gluon polarization.}
}
%*************************************************************************
\FIGURE[h]{
  \includegraphics*[width=13.5cm]{./figs/asym_pairs_paper_std_hsg.epsi}
   \caption{\label{fg:2h_asym_dp} Measured asymmetry for hadron pairs 
    produced from hydrogen (left) and deuterium (right) targets
    as a function of the minimum value of $\sum{p_{T(beam)}^2}$. 
    The uncertainties are statistical only. There is an
    overall experimental normalization  uncertainty of 5.2\% (3.9\%) for 
    hydrogen (deuterium). The curves show the Monte Carlo asymmetries for three
    different values assumed for the gluon polarization.}
}
%*************************************************************************

The curves in Figs.~\ref{fg:1h_at_asym_dp}, \ref{fg:1h_t_asym_dp}, 
and~\ref{fg:2h_asym_dp} show the asymmetries calculated by the
procedure discussed in Sect.~\ref{sec:physics}
using the values \dgg\ =  $-1$, $0$, $+1$ (from top to bottom). They 
illustrate the sensitivity of the \hermes\ data to \dgg. 
The data are close to the central curve indicating
small average values of \DGG.

%%%%%%%%%%%%%%%%%%%%%%%%%%%%%%%%%%%%%%%%%%%%%%%%%%%%%%%%
%%%%%%%%%%%%%%%%%  Extraction Procedure %%%%%%%%%%%%%%%%%
%%%%%%%%%%%%%%%%%%%%%%%%%%%%%%%%%%%%%%%%%%%%%%%%%%%%%%%%%%%%%%%%%%%%%%%%%%%
\section{Physics Model}
\label{sec:physics}
%%%%%%%%%%%%%%%%%%%%%%%%%%%%%%%%%%%%%%%%%%%%%%%%%%%%%%%%%%%%%%%%%%%%%%%%%%%
\subsection{Subprocesses}
%%%%%%%%%%%%%%%%%%%%%%%%%%%%%%%%%%%%%%%%%%%%%%%%%%%%%%%%%%%%%%%%%%%%%%%%%%%
Both the helicity-averaged and helicity-dependent cross sections include contributions 
from hard subprocesses that can be calculated using pQCD and from soft 
subprocesses such as those described by the Vector-Meson Dominance (VMD) model 
(see Fig.~\ref{fg:qcddiag}). A smooth transition from soft subprocesses to hard 
subprocesses is regulated by a set of cutoff parameters
(for details \cite{hermes:liebing, PYTHIA6.2, Sjostrand:2001yb,lund:friberg}).
The measured asymmetry is the weighted sum of the asymmetries
of all subprocesses. When it is impossible to reliably separate
the subprocesses experimentally,  as in fixed-target experiments, 
the fractions of events originating from the different subprocesses
must be modeled. 
In the analysis described in this paper, this is done using
the {\bf spin independent} Monte Carlo (MC) program 
\Pythia\ 6.2~\cite{PYTHIA6.2,Sjostrand:2001yb,lund:friberg}. 

The various subprocesses are classified in terms of the model 
used in \Pythia. In this model, the 
wave function of the incoming photon has three 
components, a ``VMD'',  a ``direct'' and an ``anomalous'' one. 
The generic photon processes following from this decomposition 
are depicted in Fig.~\ref{fg:resolved}. 
The direct photon interacts as a point-like particle with the partons of the 
nucleon, while the VMD and anomalous components interact through their 
hadronic structure.

Figure~\ref{fg:resolved}b shows an example of a direct process. 
The direct pQCD subprocesses studied in this analysis are the \ordera{0} DIS 
process (Fig.~\ref{fg:qcddiag}a), the \ordera{1} processes PGF 
(Fig.~\ref{fg:qcddiag}b), and QCDC (Fig.~\ref{fg:qcddiag}c).

The VMD component is characterized by small-scale, non-perturbative
fluctuations of the photon into a $q \bar q$ pair existing 
long enough to evolve into a hadronic state before 
the interaction with the nucleon. This process can be described 
in the framework of the VMD model, where 
the hadronic state is treated as a vector meson  
(\textit{e.g.}, $\rho^0,\,\omega,\,\phi$) with the same quantum 
numbers as the photon. 
Higher-mass and non-resonant states are added in the  
Generalized VMD (GVMD) model.  
The (G)VMD hadronic states can undergo all the interactions 
with the nucleon allowed in hadronic physics, \textit{i.e.}, elastic 
and diffractive as well as inelastic non-diffractive reactions. 
The latter can be either soft (``low-$p_T$'') processes or
hard QCD $2\rightarrow2$ processes. A generic example of a 
VMD process is shown in Fig.~\ref{fg:resolved}a.
%*************************************************************************
\FIGURE[h]{
\includegraphics*[width=13.5cm]{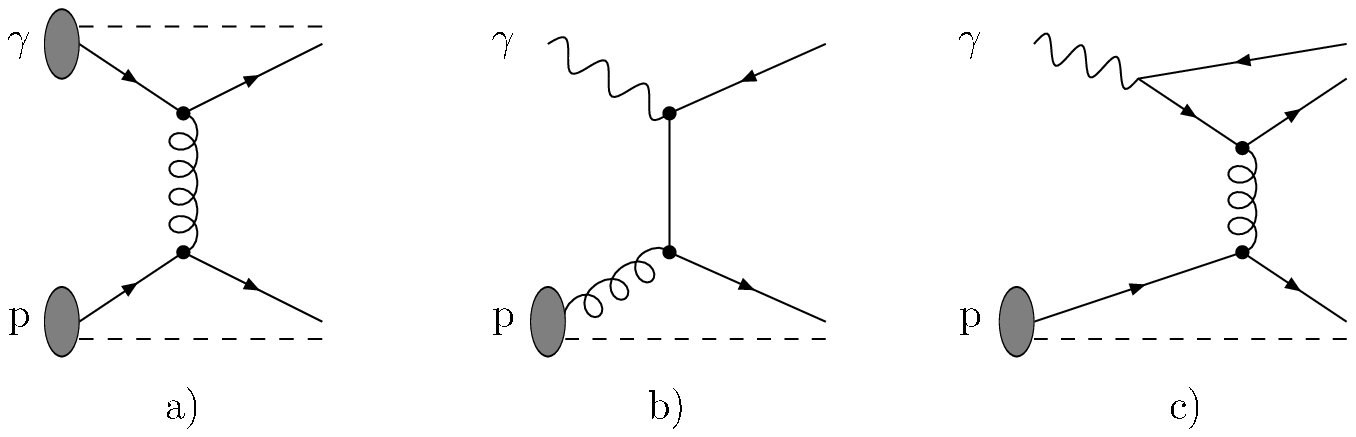}
\caption{Contributions to hard p interactions: (a) VMD, (b) direct, and 
(c) anomalous. Only the basic graphs are illustrated; additional partonic 
activity is allowed in all three processes. The presence of spectator jets 
has been indicated by dashed lines, while full lines show partons that (may) 
give rise to high-p$_t$ jets. The gray ovals represent multiparton wave 
functions. Anomalous states are built up from a perturbatively given $q \bar q$ 
fluctuation, while VMD fluctuations allow no simple perturbative 
representation; hence the difference in the placing of the ovals
\protect \cite{lund:friberg}.}
  \label{fg:resolved} 
}
%*************************************************************************

The anomalous photon is characterized by sufficiently 
large-scale, perturbative fluctuations of the photon into a $q \bar q$ pair. 
%where the subsequent evolution of the $q \bar q$ state is described 
%by the DGLAP evolution equations. 
The allowed processes  are the same pQCD $2\rightarrow2$ 
processes as in the hard VMD case, with the difference 
that for the anomalous component the parton distributions
of the photon are relevant, whereas for the description 
of the hard VMD component those of the vector meson are used. 
Both hard VMD and anomalous components are usually referred 
to as ``resolved'' photons. Depending on whether a quark or a gluon in the 
nucleon is struck by a resolved photon the corresponding hard subprocesses 
are labeled by a `q' or `g' in this paper. A generic example of an 
anomalous process is shown in Fig.~\ref{fg:resolved}c.
The resolved-photon processes are of \ordera{1} with
a hidden $1/\alpha_s(\mu^2)$ term in the evolution of the photon's 
parton distributions canceling the additional vertex~\cite{th:Drees}.

For hard subprocesses the nucleon is described by helicity-averaged 
(helicity-dependent) PDFs, which are the average (difference) of
the number densities of partons of type f whose spins are 
aligned, $f^+$, those whose spins are
anti-aligned, $f^-$, with respect to the nucleon spin:
$f(x,\mu^2)$=$f^+(x,\mu^2)$ $+$ $f^-(x,\mu^2)$ 
($\Delta f(x,\mu^2)$=$f^+(x,\mu^2)$ $-$ $f^-(x,\mu^2)$), 
where $f$= $u$, $d$, $s$, or $g$.
The integral over $x$,
$\Delta f(\mu^2) = \int^1_0 dx \Delta f(x,\mu^2)$, gives the total spin
contribution of the respective partons to the nucleon spin,
 as used in Eq.~\ref{eq:spintot}.  The hard part of the single-inclusive 
differential helicity-dependent 
cross section for the process $\gamma^*p \rightarrow hX$ can be
expressed as an integral over the parton distribution functions, 
the hard partonic cross sections for the subprocesses 
$ab \rightarrow cX$, and the fragmentation functions.  It 
can be written schematically as
%*************************************************************************
\begin{eqnarray}
   d\Delta \sigma^{\gamma^* p\rightarrow hX}&=&
   \sum_{a,b,c=q,\bar q,g}
   \int\, dx_adx_bdz_c \Delta f_a^{\gamma^*}
   (x_a,\mu^2) \Delta f_b^N(x_b,\mu^2)\nonumber\\
   &&
   \times d\Delta\hat{\sigma}^{a b\rightarrow cX}
   (\hat s,\hat t,\mu^2, Q^2)D_c^h(z_c,\mu^2) \ ,
   \label{eq:pQCD}
\end{eqnarray}
%*************************************************************************
and correspondingly for the helicity-averaged cross section and distributions. 
Here $x_b$ is the fraction of the nucleon momentum carried 
by parton $b$ and $f_b^N(x_b,\mu^2)$ ($\Delta f_b^N(x_b,\mu^2)$) is the corresponding
%(helicity-dependent) 
nucleon PDF. Similarly $x_a$ is the fraction 
of the photon momentum carried by parton $a$, and  
$f_a^{\gamma^*}(x_a,\mu^2)$ ($\Delta f_a^{\gamma^*}(x_a,\mu^2)$) is the corresponding 
%(helicity-dependent) 
photon PDF. For the direct-photon processes $a$ equals $\gamma^*$ and
$f_a^{\gamma^*}(x_a,\mu^2)$ ($\Delta f_a^{\gamma^*}(x_a,\mu^2)$) reduces to 
$\delta(1-x_a)$.  The fragmentation function $D_c^h(z_c,\mu^2)$ describes the 
hadronization of a parton $c$ into a hadron $h$ with a momentum 
$p_h=z_cp_c$. The hard partonic 
%(helicity-dependent) 
cross section $d\hat{\sigma}^{a b\rightarrow cX}(\hat s,\hat t,\mu^2, Q^2)$  
($d\Delta \hat{\sigma}^{a b\rightarrow cX}(\hat s,\hat t,\mu^2, Q^2)$) depends 
on the subprocess kinematics, the renormalization and factorization scales, 
and on $Q^2$ in case of the direct-photon processes.
Here, $\hat{s}$ and $\hat{t}$ are the Mandelstam variables for 
the partonic interaction, which are related to $x_a$ and $x_b$. More
information on the kinematic variables is given in
table~\ref{tb:kin}. The cross section for hadron pairs 
$d \sigma^{\gamma^* p\rightarrow h_1h_2X}$ ($d\Delta \sigma^{\gamma^* p\rightarrow h_1h_2X}$) 
can be obtained analogously to Eq.~\ref{eq:pQCD}.

The cross sections and asymmetries of the soft VMD interactions 
can only be modeled phenomenologically. The \Pythia\ model 
incorporates the total $\gamma p$ and hadronic cross section 
parameterizations of  Donnachie and Landshoff~\cite{th:Don92} together with 
quark counting rules~\cite{Sch93a,Sch93n}. 
This model successfully describes the measured total, elastic, 
and diffractive cross sections over a wide energy range. The non-diffractive 
cross section is modeled in \Pythia\ as the difference of the 
total cross section and the summed elastic and diffractive cross sections; 
the corresponding subprocess is called ``low-$p_T$''. 
The \Pythia\ model provides a smooth transition from real to  
virtual photons and is applicable from very small to large values of 
$Q^2$. 
It uses a number of cutoff, scale, and suppression parameters 
together with several possible prescriptions on how to use them
to select the underlying subprocess of an event. The default 
prescriptions and the cutoff and scale parameters were 
developed and tuned to match high energy data. 
In this application to the lower energy of \hermes\, the influence of 
various prescriptions and parameter values has been carefully studied 
(see Sects.~\ref{sec:pythia} and ~\ref{sec:SysErr}). 

%*************************************************************************
\TABLE[h]{
 \caption{Description of the subprocesses used in this paper. Columns from 
  left to right: subprocess, \Pythia\ subprocess number, classification as 
  signal or background, description, and name used in this paper. A vector 
  meson is denoted by $V$.}
\label{tb:subproc}
\begin{tabular}{|l|c|c|c|c|} \hline
Subprocess & \# & Class  & Description &Name\\
\hline
  \multicolumn{5}{|c|}{soft VMD} \\
\hline
$VN\rightarrow VN$ & 91  &background&   elastic VMD &exclusive VMD\\
$VN\rightarrow VX$ & 92  &background&   single-diffractive VMD &\\
$VN\rightarrow XN$ & 93  &background&  single-diffractive VMD&\\
$VN\rightarrow XX$ & 94  &background&  double-diffractive VMD&\\
\hline
$VN\rightarrow X$  & 95 &background &  soft non-diffractive VMD&low-$p_T$\\
\hline
  \multicolumn{5}{|c|}{RESOLVED  (hard VMD and anomalous)}\\
\hline
$qq\rightarrow qq$&11 &background& QCD $2\rightarrow2$&QCD $2\rightarrow2(q)$\\
$q\bar q\rightarrow q\bar q$&12 &background& .&.\\
$q\bar q\rightarrow gg$&13 &background& .&.\\
$gq\rightarrow gq$&28 &background& .&. \\
$qg\rightarrow qg$&28 &signal&  .&QCD $2\rightarrow2(g)$\\
$gg\rightarrow q\bar q$&53 &signal &.&.\\
$gg\rightarrow gg$&68 &signal &.&.\\
\hline
  \multicolumn{5}{|c|}{DIRECT}\\
\hline
$\gamma^*q\rightarrow q$&99 &background&LO DIS&DIS\\
$\gamma_T^*q\rightarrow qg$&131 &background&(transverse) QCDC&QCDC\\
$\gamma_L^*q\rightarrow qg$&132 &background&(longitudinal) QCDC&.\\
$\gamma_T^*g\rightarrow q\bar q$ &135 &signal&(transverse) PGF&PGF\\
$\gamma_L^*g\rightarrow q\bar q$&136 &signal&(longitudinal) PGF&.\\
\hline
\end{tabular}
}
%*************************************************************************
Table \ref{tb:subproc} shows a compilation of the modeled
reactions, the corresponding \Pythia\ subprocess numbers,
their classification, description, and name used in this paper.

%%%%%%%%%%%%%%%%%%%%%%%%%%%%%%%%%%%%%%%%%%%%%%%%%%%%%%%%%%%%%%%%%%%%%%%%%%%
\subsection{Signal and Background Asymmetries}
%%%%%%%%%%%%%%%%%%%%%%%%%%%%%%%%%%%%%%%%%%%%%%%%%%%%%%%%%%%%%%%%%%%%%%%%%%%
In the simulation, the cross section is considered to arise from an {\it
incoherent} superposition of all contributing subprocess amplitudes.
The kinematic selection criteria (\textit{e.g.}, event category and 
hadron $p_T$) 
for the Monte Carlo are the same as those for the data. \Pythia\ 
events are generated independent of helicity, therefore the MC 
asymmetry $A_{MC}$ is 
calculated by weighting each selected MC generated hadron with 
the calculated event asymmetry $w_k$. The average of these weights 
is $A_i$, the asymmetry for subprocess $i$ 
\begin{equation}
A_i=\frac1{N_i}\sum_{k=1}^{N_i}w_k ,
\end{equation}
where $N_i$ is the number of entries.
The event-by-event weighting method guarantees the correct integration over the
subprocess kinematics, and all partons in the nucleon and the photon 
(where applicable).
The Monte Carlo asymmetry $A_{MC}$ is the sum of the asymmetries from 
signal ($A^{SIG}_{MC}$) and background ($A^{BG}_{MC}$) subprocesses weighted 
by their fraction of entries $R^{SIG}$ and $R^{BG}$.
It is given by
%*************************************************************************
\begin{equation}
  \label{eq:SumProc}
  A_{MC}(p_T) =R^{BG}A^{BG}_{MC}(p_T)+R^{SIG}A^{SIG}_{MC}(p_T)  = 
  \sum_{i\in BG} \hspace{-.2em} R_iA_i + \sum_{i\in sig}
\hspace{-.2em} R_iA_i,
 \end{equation}
%*************************************************************************
where $R_i$ is the fraction of entries from the subprocess $i$ 
calculated in the PYTHIA simulation. 
Background processes are all subprocesses that do not
involve a hard gluon from the initial nucleon.
These include all soft processes, the direct processes DIS and QCDC,
and all resolved pQCD processes, which involve a quark or antiquark in the
nucleon, \textit{i.e.}, QCD $2\rightarrow2(q)$. They are listed in
Tab.~\ref{tb:subproc}.
All subprocesses involving a hard gluon
of the nucleon in the initial state are considered to be signal
processes, \textit{i.e.}, PGF and the hard $2\rightarrow 2(g)$ processes.

The event-by-event weight $w$ for hard subprocesses is given by
%*************************************************************************
\begin{equation}
\label{eq:weights}
  w=\hat{a}(\hat s,\hat t,\mu^2, Q^2)\cdot
    \frac{\Delta f_a^{\gamma^*}(x_a,\mu^2)}
                {f_a^{\gamma^*}(x_a,\mu^2)}\cdot
    \frac{\Delta f_b^N(x_b,\mu^2)}{f_b^N(x_b,\mu^2)},
\end{equation}
%*************************************************************************
where $\Delta f_a^{\gamma^*}/f_a^{\gamma^*}=1$ for $x_{a}=1$,
\textit{i.e.}, direct photon processes.
The hard subprocess asymmetry is
$\hat{a}(\hat s,\hat t,\mu^2, Q^2)=\Delta\hat{\sigma}/(2\hat{\sigma})$.
The lowest order equations for important hard subprocess
asymmetries are compiled in appendix~\ref{ap:equations}.
The VMD and GVMD diffractive subprocesses may have small asymmetries at 
\hermes\ energies~\cite{th:Fraas,th:Kochelev1,th:Kochelev2,hermes:rho2001}.
The asymmetry of the low-$p_T$ process was estimated from the measured
asymmetries and found to be non-zero (see Sect.~\ref{sec:MCevents}).
In both cases, the virtual-photon depolarization factor$D(y,Q^2)$ 
(see Eq.~\ref{depolfact})
has to be applied to the weight in order to account for the transformation
of the virtual-photon nucleon asymmetry into a lepton-nucleon asymmetry. 
The asymmetry from signal subprocesses depends on the unknown \dgg\
averaged over the subprocess kinematics in the specified $p_T$ range. 
It can be written as
%*************************************************************************
\begin{eqnarray}
\label{eq:SumSig2}
\hskip -0.8cm
 \nonumber A^{SIG}_{MC}(p_T) & = &
  \frac{1}{N^{SIG}}\sum_{k=1}^{N^{SIG}}\hspace{-.2em}w_k\\
  & = &
 \left\la \hat{a}(\hat{s},\hat{t},\mu^2,Q^2)\cdot\frac{\Delta f_a^{\gamma^*}(x_a,\mu^2)}
   {f_a^{\gamma^*}(x_a,\mu^2)}\cdot\frac{\Delta
   g}{g}(x_b,\mu^2)\right\ra^{SIG}\!\!\!\!\!\!\!\!\!(p_T),
\end{eqnarray}
%*************************************************************************
where $N^{SIG}$ is the number of entries from all signal processes.
The extraction of the quantity of interest, \dgg, is based on
Eq.~\ref{eq:SumSig2} replacing the unknown asymmetry
$A^{SIG}_{MC}(p_T)$ by
%*************************************************************************
\begin{equation}
\label{eq:SumSig3}
\hskip -0.8cm
A^{SIG}(p_T) = \frac{A_{meas}(p_T) - R^{BG}A^{BG}_{MC}(p_T)}{R^{SIG}}.
\end{equation}
%*************************************************************************
In Sect.~\ref{sec:DG} methods will be described to extract \DGG\ from
the right hand side of Eq.~\ref{eq:SumSig2}.
%%%%%%%%%%%%%%%%%%%%%%%%%%%%%%%%%%%%%%%%%%%%%%%%%%%%%%%%%%%%%%%%%%%%%%%%%
\section{Monte Carlo simulation}
\label{sec:pythia}
%%%%%%%%%%%%%%%%%%%%%%%%%%%%%%%%%%%%%%%%%%%%%%%%%%%%%%%%%%%%%%%%%%%%%%%%%
The relevant subprocess cross sections have 
been modeled by the \Pythia\ Monte Carlo program, which 
uses \Jetset\ \cite{Andersson} for describing the fragmentation process. 
The standard helicity-averaged input PDFs used are 
CTEQ5L~\cite{pdf:cteq5} for the nucleon and Schuler and 
Sj{\"o}strand~\cite{Sjostrand:pdf-gam} 
for the photon. The scale $\mu^2$ of the $2\rightarrow 2$ 
subprocesses is defined to be $\mu^2=\hat{p}_T^2+{\frac{1}{2}}Q^2$ 
(also commonly referred to as $\hat{Q^2}$).
Electromagnetic radiative effects~\cite{Mo69,Akushevich:1998ft}
have been added to \Pythia\, and 
they constitute a relatively small correction for hadron production
at \hermes\ kinematics~\cite{hermes:liebing}.
Events generated by \Pythia\ are passed through a 
complete {\sc Geant 3}~\cite{Brun:1978fy} simulation of the \hermes\ 
spectrometer.

%%%%%%%%%%%%%%%%%%%%%%%%%%%%%%%%%%%%%%%
\subsection{Monte Carlo tuning}
\label{sec:MCtune}
%%%%%%%%%%%%%%%%%%%%%%%%%%%%%%%%%%%%%%%
In order to account for the relatively 
low \mbox{center-of-mass} energy of the \hermes\ experiment several 
parameters in the event generation were adjusted and the model describing
exclusive vector meson production was improved~\cite{hermes:liebing}.
This was done in the kinematic region of the tagged events because 
more kinematic variables are measured for this category than for the 
anti-tagged category.
The tuning of the fragmentation parameters~\cite{Hillenbrand:2005} was 
performed using a subsample with $p_{T(\gamma^*)}<0.8$ GeV and $Q^2>1$ GeV$^2$ 
where the DIS process (Fig.~\ref{fg:qcddiag}a) is dominant and
NLO corrections are small. 
The values of the adjusted parameters, shown in Tab.~\ref{tb:pystd} 
in appendix~\ref{sec:pythia_param}, are used for all event categories.

Figure~\ref{fg:cs_xQ} shows the measured and the simulated cross sections 
as a function of $x_B$, $Q^2$, and $z$ for the tagged category of events using
a deuterium target. Both the simulated and measured cross sections are not 
corrected for acceptance effects. These cross sections vary over more than 
three orders of magnitude. The data and MC simulation agree to within 15\%  
for $x_{B}<0.2$, where most of the data reside for the tagged event category.
Thus in this region the modified \Pythia\ 6.2 program with the adjusted 
parameters gives a good representation of the cross section at \hermes\ energies.
%***********************************************************************
\FIGURE{
  \includegraphics*[width=5.5in]{./figs/pythia_xsec_tagged_hrc.epsi}
  \caption{\label{fg:cs_xQ} Top panels: Measured cross section in the
  \hermes\ acceptance for tagged hadrons as a function of $x_{B}$ (left), 
  $Q^2$ (middle), and $z$ (right) for positive (full points) and 
  negative hadrons (open points) using a deuterium target.
  The lines show the tuned \Pythia\ 6.2 calculation. Bottom panels:
  The corresponding ratios of the \Pythia\ calculation to the 
  measured cross section.}
}
%***********************************************************************

The description of the kinematic dependences of the tuned Monte Carlo code for
the individual subprocesses must be consistent with independent 
LO pQCD calculations~\cite{th:Jager}. 
Such calculations presently exist only for inclusive $\pi^0$ production and  
only in the collinear approach, where the intrinsic transverse momentum 
$k_T$ of the partons in the nucleon and in the virtual photon, and also the 
transverse momentum $p_T^{frag}$ arising from the fragmentation process are 
set to zero.

For a comparison of \Pythia\ with these LO pQCD calculations a {\bf special} 
simulation with $k_T=0$ and $p_T^{frag}=0$ was performed, 
by replacing the string fragmentation performed by JETSET with weights
obtained from the fragmentation functions of Ref.~\cite{th:kkp}.
The resulting transverse momentum $p_{T(beam)}$ of the $\pi^0$ is 
calculated according to $p_{T(beam)} = z\cdot \hat{p}_T$.
Both this simulation and the pQCD calculation are performed
in the \hermes\ kinematics for inclusive $\pi^0$ production at 
$Q^2 < 0.01$ GeV$^2$, $0.2 < y < 0.9$, disregarding the detector acceptance. 
Figure~\ref{fg:cs_pythia_bmw} compares the resulting cross sections for 
resolved photon, QCDC, and PGF processes from the simulation and the pQCD 
calculation. In the collinear approach the DIS subprocess is not included, 
because the $p_{T(\gamma^*)}$ of the final state hadron is zero, and also for 
low $Q^2$ ($Q^2<0.01$ GeV$^2$) it does not result in a sizable $p_{T(beam)}$.

%***********************************************************************
\FIGURE{
  \includegraphics*[width=5.2in]{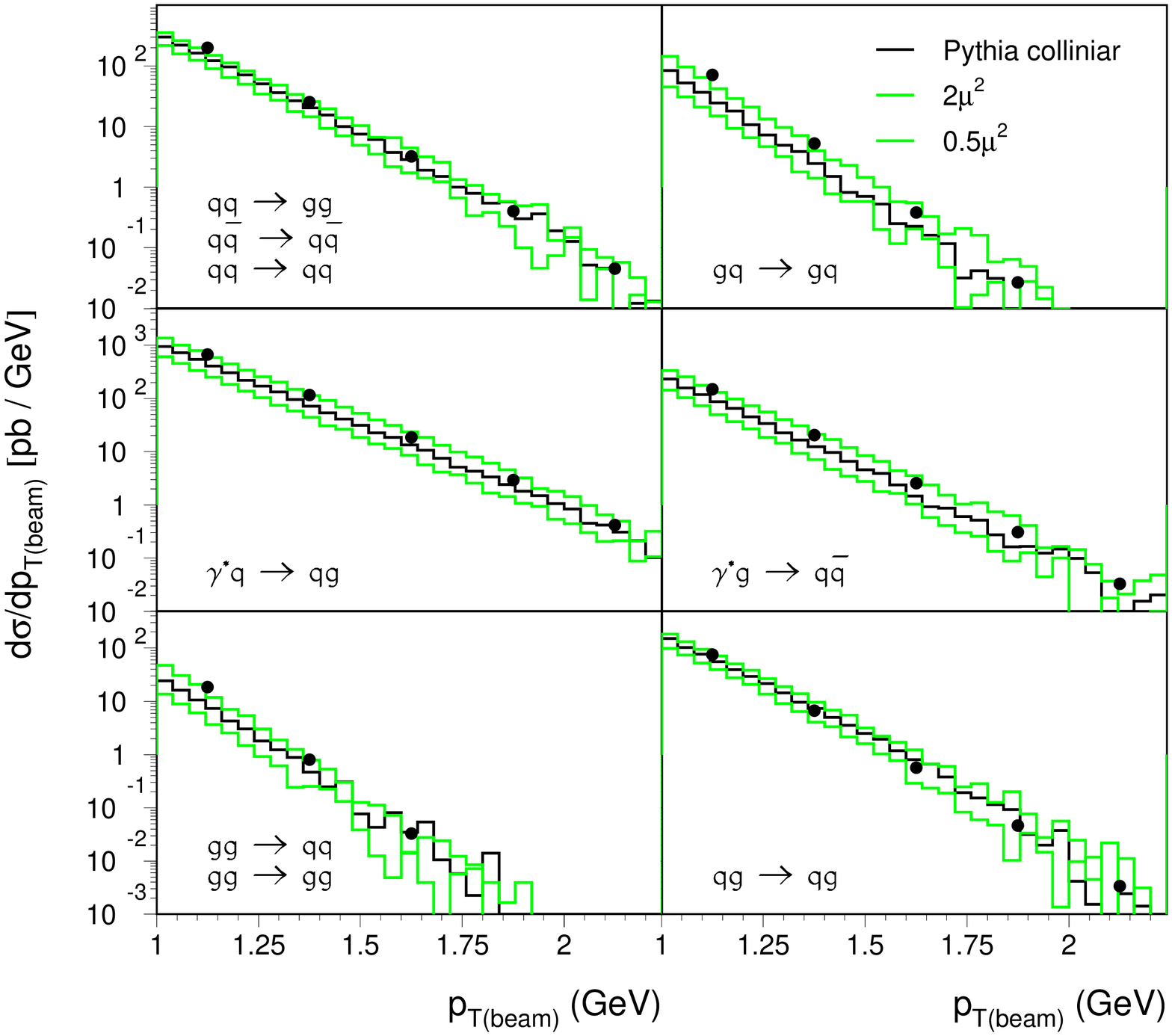}
  \caption{\label{fg:cs_pythia_bmw} Cross sections for inclusive $\pi^0$ 
   production from resolved photon,
   QCDC, and PGF processes simulated using \Pythia\ (solid lines) 
   compared to the LO pQCD calculations from
   Ref.~\cite{th:Marco} (full points). Simulation and calculation are done
   in the collinear approach at $Q^2 < 0.01$ GeV$^2$, $0.2 < y < 0.9$.  
   Green/Grey lines: subprocess
   cross sections after varying the renormalization and factorization scales
   by factors of $\frac{1}{2}$ and $2$ in the simulation.}
}
%***********************************************************************

The agreement between the simulated cross sections for the 
individual subprocesses and the calculations is well 
within the scale uncertainty ($\frac{1}{2}\mu^2, 2\mu^2$) of the 
simulation (the dashed lines in Fig.~\ref{fg:cs_pythia_bmw}). 
The LO pQCD calculations 
show a similar dependence on the variation of the renormalization 
and factorization scales ($\frac{1}{2}p_{T(beam)}, 2p_{T(beam)} $),  
see Fig.~11 in Ref.~\cite{th:Jager}.

%%%%%%%%%%%%%%%%%%%%%%%%%%%%%%%%%%%%%%%%%%%%%%%%%%%%%%%%%%%%%%%%%%%%%%%%%
\subsection{Effects of intrinsic and fragmentation transverse momenta}
\label{sec:kt}
%%%%%%%%%%%%%%%%%%%%%%%%%%%%%%%%%%%%%%%%%%%%%%%%%%%%%%%%%%%%%%%%%%%%%%%%%
While the effect of intrinsic and fragmentation transverse momenta cannot 
yet be studied in LO pQCD calculations, a \Pythia\ simulation can be used.
For the {\bf standard} simulations presented in this analysis a Gaussian 
distribution with a 0.4 GeV width is used for both $k_T$ and 
$p_T^{frag}$~\cite{Hillenbrand:2005}. 
These values are consistent with those obtained in Ref.~\cite{th:anselmino}.
Both intrinsic and fragmentation transverse momenta
alter the relationships of $\hat{p}_T^2$ to $p_{T(beam)}$, from
$p_{T(beam)} = z\cdot \hat{p}_T$ to 
$p_{T(beam)}=z(k_T+\hat{p}_T)+p_T^{frag}$, and hence the distribution of 
$\hat{p}_T^2$ and $x$.
% for a given range of $p_{T(beam)}$ ($1-2$ GeV).
This in turn influences the dependence of the cross section on $p_{T(beam)}$.
The effects on the cross section for inclusive $\pi^0$ production from
the PGF subprocess, of first adding nonzero $k_T=0.4$ GeV and secondly 
using \Jetset\ with $p_T^{frag}=0.4$ GeV are shown 
in Fig.~\ref{fg:cs_pgf_diffpt}. 
Including only $k_T$ in the simulation decreases 
$\langle \hat{p}_T^2 \rangle$ from 1.9 GeV$^2$ to 1.6 GeV$^2$ and 
$\langle x \rangle$ from 0.32 to 0.28, and 
increases the cross section by a factor of two.
Including both $k_T$ and $p_T^{frag}$ further decreases 
$\langle \hat{p}_T^2 \rangle$ to 1.1 GeV$^2$ and $\langle x \rangle$ to 0.22, 
and increases the cross section by another factor of 10.
These studies show that at fixed-target kinematics, like at \hermes,
intrinsic and fragmentation transverse momenta cannot be neglected in pQCDC
calculations. Similar conclusions were drawn in Ref.~\cite{th:HJJC}. 
Perhaps resummation techniques \cite{th:resummation}, which account for 
initial and final state radiation effects, can help to achieve more 
realistic calculations.
%***********************************************************************
\FIGURE{
  \includegraphics*[width=15cm]{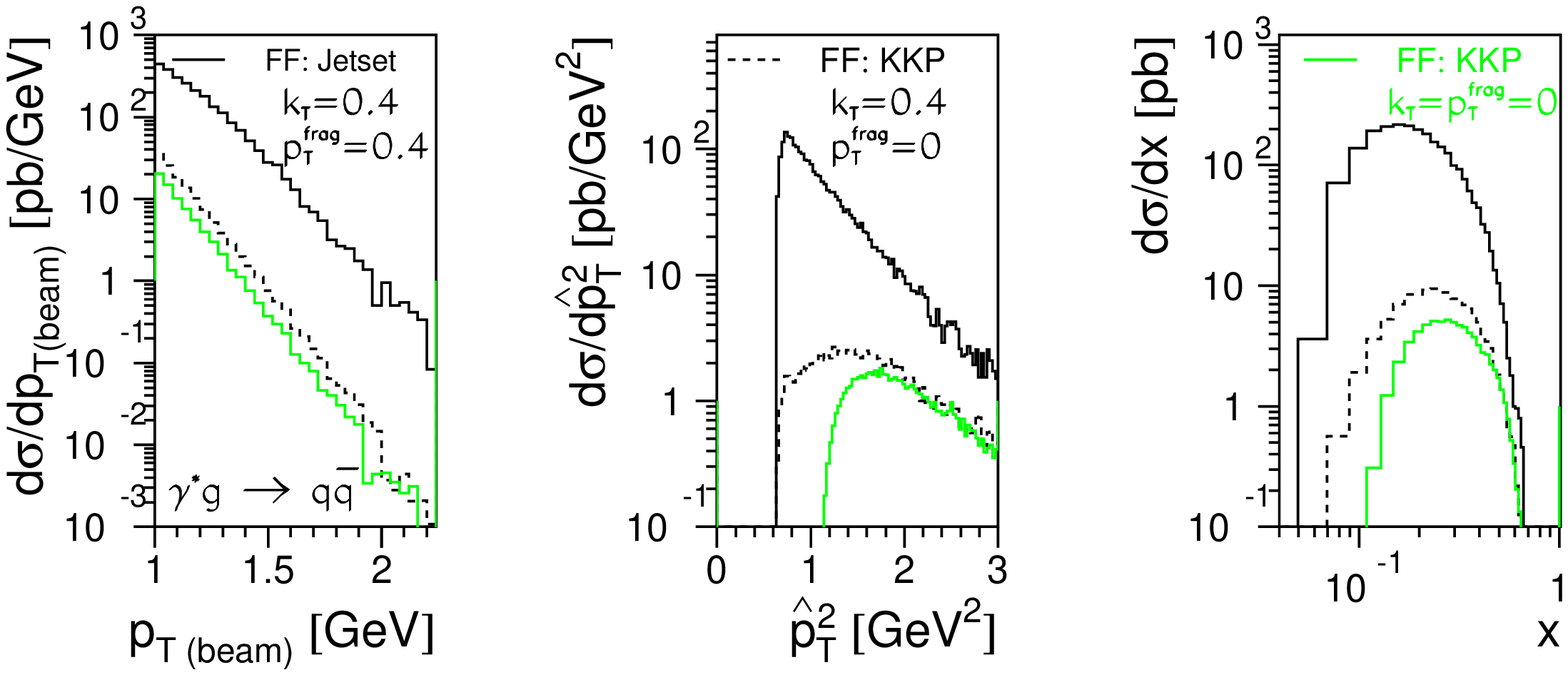}
  \caption{\label{fg:cs_pgf_diffpt}
   The simulated cross section for inclusive $\pi^0$ production from 
   the PGF subprocess vs. $p_{T(beam)}$, $\hat{p}_T^2$, and $x$  
   for $Q^2 < 0.01$ GeV$^2$, $0.2 < y < 0.9$.  
   The simulations are done using the collinear approach ((green/gray) 
   solid line), the collinear approach together with intrinsic $k_T$ for 
   the partons in nucleon and photon (dashed line), and with intrinsic 
   $k_T$ together with fragmentation transverse momenta (solid line).
   For the first two simulations fragmentation is modeled using the 
   KKP-fragmentation functions \protect \cite{th:kkp}, for the third 
   one \Jetset\ with the standard settings listed in 
   Table \protect \ref{tb:pystd} is used.}
}
%***********************************************************************

%%%%%%%%%%%%%%%%%%%%%%%%%%%%%%%%%%%%%%%%%%%%%%%%%%%%%%%%%%%%%%%%%%%%%%%%%
\subsection{Analysis of Monte Carlo events}
\label{sec:MCevents}
%%%%%%%%%%%%%%%%%%%%%%%%%%%%%%%%%%%%%%%%%%%%%%%%%%%%%%%%%%%%%%%%%%%%%%%%%
\Pythia\ events are used to calculate cross sections, individual subprocess 
fractions $R_i$ and event weights $\langle w \rangle_i$ 
within the \hermes\ acceptance.
The event weights for the pQCD processes (Eq.~\ref{eq:weights}) are obtained 
using the hard subprocess asymmetries (see appendix \ref{ap:equations}) and 
GRSV (standard scenario)~\cite{pdf:grsv2000} helicity-dependent PDFs 
in conjunction with the GRV98~\cite{pdf:grv98} helicity-averaged PDFs to calculate
$\Delta f^N/f^N$ for the nucleon.
In order to calculate $\Delta f^{\gamma}/f^{\gamma}$ for the photon 
the averages of the maximal and minimal scenarios of the 
GRS~\cite{ref:GRS-PG,ref:GRS-PG2} helicity-dependent PDFs are used in conjunction 
with the GRS~\cite{pdf:grs1999} helicity-averaged PDFs.

For elastic and diffractive VMD processes the asymmetry is set to 
zero~\cite{hermes:rho2001}.
For the low-$p_T$ process two alternative assumptions for the asymmetry
have been investigated: $A_{low\mbox{-}p_{T}}=D(y,Q^2)\cdot A_1$ and 
$A_{low\mbox{-}p_{T}}=0$ where $A_1$ is a parameterization of the 
photon-nucleon asymmetry in inclusive DIS.
% and $D$ is the so-called depolarization factor given by Eq.~\ref{depolfact}.
The resulting MC asymmetries and the measured
asymmetry are shown in Fig.~\ref{fg:apar-lowpt} (left) 
vs. $x_B$ for the tagged category and a hydrogen target. 
The corresponding deuterium data are not shown because for this target both 
assumptions are indistinguishable and match the data.
For the anti-tagged category (Fig.~\ref{fg:apar-lowpt} (right)) 
the $p_{T(beam)}$
dependence of the double-spin inclusive asymmetry $A_\parallel$ is shown 
for both targets. The model $A_{low\mbox{-}p_{T}}=D\cdot A_1$ matches the data 
better than $A_{low\mbox{-}p_{T}}=0$ in the kinematic domains 
where $R^{low\mbox{-}p_T}$ is large (low-$x_B$ for tagged
and the lowest $p_T$ for anti-tagged categories, respectively) and 
contributions of hard QCD processes are negligible.
The standard asymmetry for the low-$p_T$ process was chosen to
be $A_{low\mbox{-}p_{T}}=D\cdot A_1$, because of this agreement
and because the semi-inclusive asymmetry for all charged hadrons
is approximately equal to the measured inclusive asymmetry.
 The world data on $A_1$ have been parameterized by 
$a+x_B^b\cdot(1-e^{cx_B})$ for $x_B > 10^{-3}$, 
and extrapolated to the smaller $x_B$-values 
($\langle x_B \rangle \sim 0.0001$) typical for the anti-tagged sample.
%************************************************************************
\FIGURE{
 \includegraphics[width=11cm]{./figs/apar_lowpt_hrc.epsi}
 \caption{\label{fg:apar-lowpt} Left panels: The measured double-spin inclusive 
  asymmetry $A_\parallel$ (full points) and two MC asymmetries (solid and dashed
  lines) based on
  different assumptions for the low-$p_T$ subprocess asymmetry (top)
  and the process fractions vs. $x_B$ (bottom) for tagged events on a 
  hydrogen target. 
  Right panels: Double-spin asymmetry vs. $p_{T(beam)}$ for anti-tagged hadrons
  on a hydrogen (top) and deuterium (bottom) target.}
}
%************************************************************************

To avoid any bias from the experimental trigger to the results presented, 
the MC events received an additional weight to account for trigger 
inefficiencies, if measured and simulated cross sections are compared.
The $p_T$ dependences of the cross sections, individual subprocess
fractions $R_i$, average event weights $A_i$, and weighted asymmetries 
$R_iA_i$ for the three event categories
are shown in Figs.~\ref{fg:frac1h_photo} (anti-tagged),
\ref{fg:frac1h_trans} (tagged), and \ref{fg:2h_procfrac_sum} (pairs).

All three categories have in common that: 
\begin{itemize}
\item The cross sections span four orders of magnitude, decreasing rapidly with
   $p_T$;
\item Reasonable agreement between data and Monte Carlo is observed for 
   low transverse momenta. With increasing $p_T$ the Monte Carlo description 
   becomes worse, underestimating the data by up to a factor of four at 
   the largest $p_T$;
\item The fractions $R^{low-p_T}$ and $R^{excl.VMD}$ decrease with 
   increasing $p_T$ and the corresponding asymmetries are very small or
   zero, respectively;
\item In general the contributions from hard QCD subprocesses 
   increase with increasing $p_T$. At high $p_T$ subprocesses 
   involving quarks in the nucleon contribute less than the signal processes;
\item The asymmetries for the two signal subprocesses, QCD $2\rightarrow2(g)$ 
   and PGF
   have opposite sign. For a positive gluon polarization like that 
   of GRSV, this results in a sizable negative asymmetry for PGF, 
   and positive 
   asymmetries for the $2\rightarrow2(g)$ processes;
\item Some asymmetries and fractions depend on the charge of the hadron.
\end{itemize}
 
Even though soft effects from initial and final state radiation and additional
nonperturbative processes are taken into account in the \Pythia\ simulation, the
Monte Carlo simulation still fails to describe the cross sections 
at $p_T>1$ GeV. This
shortcoming may be explained by missing large higher order corrections to the
hard processes. These corrections have been evaluated for the next to leading
order (NLO) cross section in Ref. \cite{th:Jager}, in the collinear approach 
for $Q^2<0.01$ GeV$^2$ and $p_T^{col}=z\hat p_T>1$ GeV, for all hard processes 
(QCD $2\rightarrow2$, PGF, QCDC) contributing in this region. 
The similar kinematics of hard processes in the pQCD-calculation and the 
PYTHIA simulation allows one to approximate the effect of NLO corrections to
the Monte Carlo cross section. A $k$-factor, \textit{i.e.}, the ratio of LO
to NLO cross sections is applied as a weight to each hadron originating from a hard
process. The $k$-factors from Ref.~\cite{th:Jager} are very large
(almost 5) at $p_T\approx1$ GeV and decrease with $p_T$ to about 2.5 at
$p_T=2.4$ GeV. For the reweighting of the Monte Carlo events they have
been extrapolated down to $p_{T(beam)}=0.8$ GeV, and it was assumed that
$p_T^{col}$ can be approximated by $p_{T(beam)}$ (see the discussion in 
Sect.~\ref{sec:MCtune} about the collinear approximation). 
The results shown in the cross section ratio of Fig. \ref{fg:frac1h_photo} 
indicate that the inclusion of NLO effects to the
Monte Carlo could significantly improve the description of the cross section.
Effects of similar size may exist for the other categories, but NLO
calculations for those are not yet available. 
The $k$-factors for the asymmetry have also been calculated
in \cite{th:Jager} and are approximately 2 in the experimental range.
Unfortunately it is not possible
to consistently take into account $k$-factors in the extraction of \dgg,
therefore the result will essentially be a LO result subject to potentially
large NLO corrections.
%************************************************************************
\FIGURE{
 \includegraphics[width=14cm]{./figs/photo_paper_2004c_std_hrc.epsi}
 \caption{\label{fg:frac1h_photo}
  The cross sections and subprocess contributions, in the \hermes\ acceptance
  as a function of $p_{T(beam)}$ for the anti-tagged category of events and a
  deuterium target (left: positively charged hadrons, right: negatively
  charged hadrons).
  Top: The measured cross section and that generated by \Pythia .
  Second row: The ratio of these two cross sections.
  Also shown is the effect of the $k$-factor based on Ref.~\cite{th:Jager}
  (see text).  Third row: The subprocess fractions from \Pythia.
  Bottom two rows: The asymmetries and the asymmetries weighted with the
  subprocess fractions for each subprocess using Refs.~\cite{pdf:grsv2000} 
  and~\cite{pdf:grv98} for the gluon PDFs.}
}
%************************************************************************
%************************************************************************
\FIGURE{
 \includegraphics*[width=14.cm,clip]{./figs/semi_paper_2004c_std_hrc.epsi}
 \caption{\label{fg:frac1h_trans}
  The cross sections and subprocess contributions, in the \hermes\ acceptance,
  as a function of $p_{T(\gamma^*)}$ for the tagged category of events and a
  deuterium target (left: positively charged hadrons, right: negatively
  charged hadrons).
  Top: The measured cross section and that generated by \Pythia .
  Second row: The ratio of these two cross sections.
  Third row: The subprocess fractions from \Pythia.
  Bottom two rows: The asymmetries and the asymmetries weighted with the
  subprocess fractions for each subprocess using Refs.~\cite{pdf:grsv2000} 
  and~\cite{pdf:grv98} for the gluon PDFs.}
}
%************************************************************************
%************************************************************************
\FIGURE{
 \includegraphics*[width=14cm]{./figs/pairs_paper_2004c_std_hsg.epsi}
 \caption{\label{fg:2h_procfrac_sum}
  The cross sections and subprocess contributions, in the \hermes\ acceptance,
  as a function of the minimum value of $\sum{p_{T(beam)}^2}$ for the
  production of inclusive hadron pairs on a deuterium target.
  Top: The measured cross section and that generated by \Pythia.
  Second row: The ratio of these two cross sections.
  Third row: The subprocess fractions from \Pythia.
  Bottom two rows: The asymmetries and the asymmetries weighted with the  
  subprocess fractions for each subprocess using Refs.~\cite{pdf:grsv2000}   
  and~\cite{pdf:grv98} for the gluon PDFs.}
}
\clearpage
%*************************************************************************

For the anti-tagged category the LO DIS fraction dominates 
the yield of positive hadrons at high $p_{T(beam)}$. 
This is due to the subsample of events with the positron having a large 
scattering angle and missing the \hermes\ acceptance.
The subprocess fractions for LO DIS and QCDC are larger for positive hadrons
because of  $u$-quark dominance. Both signal subprocesses   
contribute approximately 20\% to  the cross section at high $p_{T(beam)}$. 
The pairs category has a larger signal fraction than the other categories, 
but a much smaller number of events.
The mixture of the background processes and their contribution to 
the background asymmetry is different for each category.

%%%%%%%%%%%%%%%%%%%%%%%%%%%%%%%%%%%%%%%%%%%%%%%%%%%%%%%%%%%%%%%%
%%%%%%%% section{Determination of the Gluon Polarization} %%%%%%
%%%%%%%%%%%%%%%%%%%%%%%%%%%%%%%%%%%%%%%%%%%%%%%%%%%%%%%%%%%%%%%%
%%%%%%%%%%%%%%%%%%%%%%%%%%%%%%%%%%%%%%%%%%%%%%%%%%%%%%%%%%%%%%%%%%%%%%%%%%%%%%
\section{Determination of the gluon polarization}
\label{sec:DG}
%%%%%%%%%%%%%%%%%%%%%%%%%%%%%%%%%%%%%%%%%%%%%%%%%%%%%%%%%%%%%%%%%%%%%%%%%%%%%%
\subsection{Kinematic considerations and requirements} 
%%%%%%%%%%%%%%%%%%%%%%%%%%%%%%%%%%%%%%%%%%%%%%%%%%%%%%%%%%%%%%%%%%%%%%%%%%%%%%

The average value of \DGG\ in a $p_T$ range is determined directly from
Eq.~\ref{eq:dgextract} (see Sect.~\ref{sec:GluonPolPT}). However, as shown in
Fig.~\ref{fg:pT_xG}, there is a large range of $x$ spanned by the
data for each $p_T$ range. In order to circumvent this difficulty,
the value of \DGG\ and the appropriate value of $x$ is
determined through a minimization procedure using a functional form for
\dggx\ (see Sect.~\ref{sec:GluonPolx}).
The scale dependence of \DGG\ is neglected because almost
all pQCD models are monotonic and vary slowly as a function of $\mu^2$ over the
relatively small relevant range.
%*************************************************************************
\FIGURE{
\includegraphics*[width=12.2cm]{./figs/pythia_xdistribution_paper.epsi}
\caption{\label{fg:pT_xG}
 The range of generated $x$ for different values of $p_{T(beam)}$ calculated 
 by \Pythia\ for all signal processes, for the anti-tagged category of events 
 and a deuterium target. }
}
%*************************************************************************
In order to optimize the accuracy of $\frac{\Delta g}g$ the following criteria 
that maximize the sensitivity of the MC asymmetry
to $\frac{\Delta g}g$, are applied to the individual data samples:

%*************************************************************************
\begin{tabular}{llclccc}
$\bullet$ & 1.0 GeV & $<$ & $p_{T(\gamma^*)}$ & $<$ & 2.0 GeV & (tagged);\nonumber\\
$\bullet$ & 1.0 GeV & $<$ & $p_{T(beam)}$ & $<$ & 2.5 GeV & (anti-tagged);\nonumber\\
$\bullet$ & 2.0 GeV$^2$ & $<$ & $\sum{p_T^2}$ & & &  (pairs).\nonumber
\end{tabular}\\
%*************************************************************************

\noindent These requirements balance the statistical accuracy of the measured 
asymmetries (decreasing with $p_T$, as shown in 
Figs.~\ref{fg:1h_at_asym_dp}-\ref{fg:2h_asym_dp}) against 
the signal process fractions (increasing with $p_T$, as shown 
in Figs.~\ref{fg:frac1h_photo}-\ref{fg:2h_procfrac_sum}). 
For the events within these limits it is observed that:
\begin{itemize}
\item The \Pythia\ simulations displayed in Fig.~\ref{fg:pthard} show a 
   strong correlation between the hard scattering transverse 
   momentum $\langle \hat{p}_T^2 \rangle$ of 
   the signal subprocesses and the measured hadronic $p_T$ ($\sum{p_T^2}$);
\item For larger values of $p_T$, there is greater sensitivity to the hard 
  processes involving the gluon
  (see Figs.~\ref{fg:1h_at_asym_dp}-\ref{fg:2h_asym_dp}), which leads 
  to reduced systematic uncertainties due to corrections for background 
  asymmetries.
\end{itemize}

%*************************************************************************
\FIGURE{
 \includegraphics*[width=13cm]{./figs/pythia_ptq2hat_std.epsi}
 \caption{\label{fg:pthard} 
  The correlation of the average hard scattering $\langle \hat p_T^2 \rangle$ 
  of all signal subprocesses as calculated by \Pythia\ with the hadron 
  $p_T$ for inclusive hadrons as calculated for the experimental data 
  for the deuterium target.
  Left: tagged category; Center: anti-tagged category; 
  Right: hadron pairs category.
  The dotted line goes along $\langle \hat{p}_T^2 \rangle = p_T^2(h)$ 
  ($\langle \hat{p}_T^2 \rangle = \sum{p_T^2}/2$) and the vertical dashed line
  shows the minimum $p_T$ ($\sum{p_T^2}$) used for the analysis. } 
}
%*************************************************************************

The gluon polarization \DGG\ is determined using Eqs.~\ref{eq:SumProc},
\ref{eq:SumSig2}, and~\ref{eq:SumSig3}. 
The anti-tagged category has sufficient statistics to allow extraction of 
$\frac{\Delta g}g$ in four $p_{T(beam)}$ bins 
(1.0 - 1.2 - 1.5 - 1.8 - 2.5 GeV), which are obtained by combining 
the bins shown in Fig.~\ref{fg:1h_at_asym_dp} and 
table~\ref{tb:Anti-tagged}. The other categories are
represented by a single range in $p_T$.

%%%%%%%%%%%%%%%%%%%%%%%%%%%%%%%%%%%%%%%%%%%%%%%%%%%%%%%%%%%%%%%%%%%%%%%%%%%%
\subsection{$p_T$ dependence of \DGG}
\label{sec:GluonPolPT}
%%%%%%%%%%%%%%%%%%%%%%%%%%%%%%%%%%%%%%%%%%%%%%%%%%%%%%%%%%%%%%%%%%%%%%%%%%%%
If the dependence of $\frac{\Delta g}g(x,\mu^2)$ on $x$ and $\mu^2$ is weak 
in the limited kinematic range of the experiment, \dggx\ can be factored from 
the r.h.s of Eqs.~\ref{eq:SumSig2}, so that together with Eqs. \ref{eq:SumSig3}
we obtain for
the gluon polarization averaged over the covered $x$ and $\mu^2$ ranges 

%************************************************************************
\begin{equation}
  \label{eq:dgextract}
  \langle \frac{\Delta g}{g} \rangle (p_T) \equiv  \\
  \frac{A_{meas}(p_T)- R^{BG}A^{BG}_{MC}(p_T)}
  {R^{SIG}(p_T)\left\la \hat a(\hat s, \hat t, \mu^2, 
   Q^2)\frac{\Delta f_a^{\gamma^*}(x_a,\mu^2)} 
  {f_a^{\gamma^*}(x_a,\mu^2)}\right\ra^{SIG}\!\!\!\!\!\!\!\!\!(p_T)},
\end{equation}
%*************************************************************************
where the subprocess fractions and kinematics are determined 
using \Pythia.  As is shown in Fig.~\ref{fg:pT_xG}, different ranges in $p_T$ 
correspond to different ranges and distributions in $x$. 
It is intrinsic to this method that there is no knowledge on the dependence of 
$\frac{\Delta g}g$ on $x$, therefore no meaningful value of the average $x$
can be determined by this method, which nevertheless can be used as a 
consistency check between the different independent data sets.

The results for different event categories, targets and
hadron charges are listed in table~\ref{tb:delG_consistency} and shown in 
Fig.~\ref{fg:delG_consistency} as a function of $p_T$. 
The results for the pairs category are displayed at the average
$\sqrt{\sum{p_T^2}/2}$, and those for the tagged category at the 
average $p_{T(\gamma^*)}$. 
%For the tagged category the average $\hat{Q}^2$ and $x$ values are 1.91 
%GeV$^2$ and 0.21, respectively. For the pairs category the corresponding 
%values are 1.45 GeV$^2$ and 0.15.
Each of these data sets has a somewhat
different mixture of background and signal processes as a function of
$p_T$, as seen in Figs.~\ref{fg:frac1h_photo}-\ref{fg:2h_procfrac_sum}.
The measured values of \dggp\ should be equal for both targets and both 
hadron charges because of the same range in $x$ and $\mu^2$.
The values shown in  Fig.~\ref{fg:delG_consistency} within 
each category and for each  $p_{T(beam)}$ bin indeed agree in general 
within the statistical uncertainties. 
This is a strong indication that \Pythia\ provides a consistent
description of the underlying physics.
The systematic charge dependence is accounted for by assigning a systematic 
uncertainty to the value of the $p_T^{frag}$ (\Pythia\ parameter PARJ(21)).

In \cite{hermes:old-glue} the kinematic selections for the hadron pairs used to calculate
the asymmetry to extract \dggx\ was $p_T^{h_1} > 1.5$ GeV and $p_T^{h_2} > 0.8$ GeV.
These events are mostly contained in the event sample used to calculate the asymmetry 
in the left panel of fig.~\ref{fg:2h_asym_dp} if 
$\sum{(p_{T(beam)}^2)}_{min} = p_{T(beam)}^{h1})^2 + (p_{T(beam)}^{h2})^2 > 3.0$ GeV$^2$ is 
required. 
The asymmetry for hadron pairs with $\sum{(p_{T(beam)}^2)}_{min} > 3.0$ GeV$^2$ 
presented here is statistically consistent with the 
average asymmetry for $p_T^{h_1} > 1.5$ GeV and $p_T^{h_2} > 0.8$ GeV from \cite{hermes:old-glue}. 
The difference for \dggp\ obtained for the inclusive pairs of hadrons 
in this paper compared to the result presented in \cite{hermes:old-glue} can
be explained by the different treatment of the underlying signal and 
background subprocesses contributing to the asymmetry and the difference in
kinematic selections of the hadron pairs used calculating the asymmetry.
The model presented in \cite{hermes:old-glue} used only 2 subprocesses 
(PGF and QCDC) to describe the measured negative asymmetry for the proton
target. For the determination of the subprocess fractions also the VMD
process was considered, which was treated to have no subprocess asymmetry,
which is consistent with the model used in this paper.
The resulting subprocess fraction for PGF in \cite{hermes:old-glue} is bigger
than from the model presented in this paper.
This combined with the positive asymmetry for the QCDC subprocess leads to the
sizable positive gluon polarization reported in \cite{hermes:old-glue} 
(Note: $\hat a(\hat s, \hat t, \mu^2, Q^2)$
is negative for PGF in the probed kinematics).

%The gluon polarization obtained by averaging the results from the deuterium 
%target for the anti-tagged category over the whole $p_T$-range 
%1.0 GeV $< p_{T(beam)} <$ 2.5 GeV is \\ 
%\centerline{{\bf $\la \frac{\Delta g}{g}\ra =$
%$0.055 \pm 0.033 (stat) \pm 0.010 (sys-exp)^{+ 0.124}_{-0.078} (sys-models)$}}
%at $\la p_{T(beam)}\ra = 1.20$ GeV and a scale $\mu^2$ = 1.35 GeV$^2$.

%*************************************************************************
\FIGURE{
\includegraphics*[width=13cm]{./figs/deltag_paper_hrc.epsi}
\caption{\label{fg:delG_consistency} The value of \dggpfact\ determined
   in the anti-tagged category for protons (top) and deuterons (bottom)
   and positive (full points) and negative (open points) hadrons as a
   function of $p_T$. Also shown are the values for the tagged (squares)
   and pairs (triangle) category at their average respective $p_T$.
   The uncertainties shown are statistical only.}
}
%*************************************************************************
%*********************************************************************
\TABLE{
\caption{\label{tb:delG_consistency}Results for \dggp\ for the
 three categories of events, both targets and hadron charges.
 Only statistical uncertainties are shown.}

\begin{tabular}{|c||r|r||r|r|} \hline
       & \multicolumn{2}{c||}{Proton} & \multicolumn{2}{c|}{Deuteron} \\ \hline
 $\langle p_T \rangle$ (GeV) 
  &\multicolumn{1}{c|}{ h$^+$}&\multicolumn{1}{c||}{h$^-$}
  &\multicolumn{1}{c|}{ h$^+$} &\multicolumn{1}{c|}{ h$^-$} \\ \hline
      \multicolumn{5}{|c|}{anti-tagged} \\ \hline
1.11  & $-0.076 \pm 0.150$ &  $0.201 \pm 0.162$
                     & $-0.063 \pm 0.096$ & $ 0.125 \pm 0.096$ \\ \hline
1.30  &  $0.011 \pm 0.120$ &  $0.125 \pm 0.103$
                     & $-0.005 \pm 0.073$ &  $0.080 \pm 0.059$ \\ \hline
1.60  &  $0.116 \pm 0.195$ &  $0.619 \pm 0.174$
                     & $-0.087 \pm 0.119$ &  $0.149 \pm 0.093$ \\ \hline
1.98  &  $0.722 \pm 0.563$ &  $0.154 \pm 0.289$
                     &  $0.865 \pm 0.297$ &  $0.446 \pm 0.178$ \\ \hline
       \multicolumn{5}{|c|}{tagged} \\ \hline
1.16  & $-0.373 \pm 0.293$ & $-0.363 \pm 0.302$
                     & $-0.372 \pm 0.191$ & $0.119 \pm 0.174$ \\ \hline
       \multicolumn{5}{|c|}{pairs} \\ \hline
1.10  & \multicolumn{2}{c||}{$-0.079 \pm 0.196$}
                     & \multicolumn{2}{c|}{$0.282 \pm 0.122$} \\ \hline
\end{tabular}
}
%*************************************************************************

%%%%%%%%%%%%%%%%%%%%%%%%%%%%%%%%%%%%%%%%%%%%%%%%%%%%%%%%%%%%%%%%%%%%%%%%%%%%
\subsection{$x$ dependence of \DGG}
\label{sec:GluonPolx}
%%%%%%%%%%%%%%%%%%%%%%%%%%%%%%%%%%%%%%%%%%%%%%%%%%%%%%%%%%%%%%%%%%%%%%%%%%%%
As there is no assumption-free method to determine the average $x$, 
various functional forms for \dggx\ with free parameters were investigated
for extracting \dggx\ from Eqs.~\ref{eq:SumSig2} and~\ref{eq:SumSig3}. 
Assuming a functional form it is possible to convert the $p_T$ dependence of 
the asymmetry into a value of \dggx\ at an average $x$.
In contrast to the method described above, this method works for
%can accomodate 
stronger $x$ dependences of $\frac{\Delta g}{g}$. 

For a given functional form and parameter set for \dggx\, $
A^{SIG}_{MC}(p_T)$ can be 
calculated using Eq.~\ref{eq:SumSig2}. The best parameter set is 
obtained by minimizing the quantity
%*************************************************************************
\begin{equation}
  \chi^2\equiv(\Delta \vec{A})^TC^A\Delta \vec{A}\label{chi2},
\end{equation}
%*************************************************************************
where $\Delta \vec{A}$ is a vector containing the 
difference between the measured and the calculated asymmetries for 
each bin in $p_T$
%*************************************************************************
\begin{equation}
\Delta \vec{A}=\vec A_{meas}-(\vec R^{BG}A_{MC}^{BG}+\vec R^{SIG}A^{SIG}_{MC}).
\end{equation}
%*************************************************************************
The matrix $C^A$ in Eq.~\ref{chi2} is the covariance matrix including the
statistical uncertainties of the data and MC asymmetries.

A scan over an appropriately large parameter space is performed in  
order to find the parameters of the function describing \dggx\
that minimize $\chi^2$. Their covariance matrix $C^F$ is determined from 
the distribution of probabilities $P_k$ at each scan point $k$:
%*************************************************************************
\begin{equation}
C_{ij}^F=\frac{\sum_k\left(\theta_{ik}-\theta_{i}^{max}\right)\left 
(\theta_{jk}-\theta_{j}^{max}\right)P_k}{\sum_k P_k}.
\end{equation}
%*************************************************************************
In this expression, $\theta_{ik}$ is the value of parameter $i$ at  
point $k$, while $\theta_i^{max}$ is the value of parameter $i$ with 
the maximum probability. The probabilities $P_k$ can be evaluated from the 
$\chi^2$ cumulative distribution function. 
The advantage of this scan procedure is that it ensures finding
the {\it global} minimum and enables the determination of the average $x$
of the measurement using the extracted shape of $\frac{\Delta g}{g}$.
%The uncertainty on \dggx\ determined by the uncertainty on the parameter(s) is
%propagated into an uncertainty on \dggpfunc\ using Eq.~\ref{eg:dgg_meth2}.

This determination of \dggx\ can be done only for the anti-tagged
category because of the necessity of having several bins in $p_T$.
In order to satisfy the fundamental requirement for \dggx\ to vanish at $x=0$
the functions are required to behave asymptotically as
$\frac{\Delta g}g(x)\rightarrow x$ as $x\rightarrow 0$.
In addition, 
$\displaystyle{\lim_{x\rightarrow 1}}\frac{\Delta g}g(x)\rightarrow 1$ 
was required \cite{Brodsky:95}. Omitting this constraint does not significantly 
change the results.
The small number of $p_{T(beam)}$ bins available limits the choice of the 
functional forms to those with no more than two free parameters.
Several functional forms were studied, and the following two selected:
\begin{itemize}
  \item[ ] fct.~1:~ $x(1+p_1(1-x)^2)$,  
  \item[ ] fct.~2:~ $x(1+p_1(1-x)^2 +p_2(1-x)^3)$.
\end{itemize}
The parameters are restricted such that the LO positivity constraint: 
$|\frac{\Delta g}g(x)|<1$ is satisfied.
%
%*************************************************************************
\FIGURE{
 \includegraphics*[width=5.0in]{./figs/mc_data_asymmetries_tracked.epsi}
 \caption{\label{fg:MII_Asyms} Measured asymmetries with
  statistical uncertainties in four $p_T$ bins for the anti-tagged category 
  and a deuterium target, compared to calculated asymmetries using the two 
  functions.}
}
%*************************************************************************
Figure~\ref{fg:MII_Asyms} compares the $p_T$ dependence of the measured 
asymmetry with the asymmetry calculated using the functional forms 
fitted to the measured asymmetries using 
Eqs.~\ref{eq:SumProc} and~\ref{eq:SumSig2}. The $\chi^2$ per degree of 
freedom is large for both functions because of the discrepancy 
between the measured and calculated asymmetries in the highest $p_T$ bin.
No functional form was found that also accommodates the fourth data point
within the statistical uncertainty. Systematic uncertainties of the 
Monte Carlo simulation (see Sect.~\ref{sec:SysErr}) have not been used in 
this minimization; including them would reduce the $\chi^2$ value 
significantly. 

%*************************************************************************
\FIGURE{
  \includegraphics*[width=4.5in]{./figs/deltaGx2fct_final.epsi}
  \caption{\label{fg:MII_funcs} Functional forms used
  with the values and statistical uncertainty bands from the fits. Light shaded
  area: the total $x$ range spanned by the data
  (see Fig.~\ref{fg:pT_xG}); dark shaded area: the range in $x$ where
  the preponderance of the data lies.}
}
%*************************************************************************
Figure~\ref{fg:MII_funcs} shows the two functional forms of 
$\frac{\Delta g}g (x)$ and their statistical uncertainties. 
The parameter value and  uncertainties for fct.~1 are given in
table~\ref{tb:syser_anti_d_II}.
The light shaded area represents the full $x$ range spanned by the 
data,  $0.07<x<0.7$ (see Fig.~\ref{fg:pT_xG}). 
The dark shaded area represents the range of $x$ 
spanned by preponderance of the data as seen in Fig.~\ref{fg:pT_xG}.
Although there are considerable differences in the \dggx\ functional forms 
over the full $x$ range, the resulting Monte Carlo asymmetries
are not very different, as can be seen from Fig.~\ref{fg:MII_Asyms}.
From the behavior of the measured asymmetries together with the variation of 
the $x$-distribution 
(see Fig.~\ref{fg:pT_xG}) with $p_{T(beam)}$ it can be seen that any smooth 
function that describes the data leads to \dggx\ for $x<0.2$
either small and positive or slightly negative, and significantly positive at 
larger $x$. 
However, no function with so few parameters is able to describe the sudden 
change of \dggx\ at $x\approx0.2$ required to match the measured asymmetry 
in the largest $p_{T(beam)}$ bin.
The average $\frac{\Delta g}{g}$ is determined using the resulting $\frac{\Delta g}{g}(x)$

%*************************************************************************
\begin{equation}
\label{eg:dgg_meth2}
\langle \frac{\Delta g}{g} \rangle  \equiv
\frac{\Delta g}{g}(\langle x\rangle) = 
\frac{\displaystyle{\sum_{k=1}^{N^{SIG}}} \hat{a}_k(\hat s, \hat t,\mu^2, Q^2)
\frac{\Delta f_a^{\gamma^*}(x^k_a,\mu^2)}{f_a^{\gamma^*}(x^k_a,\mu^2)}\frac{\Delta g}{g}(x^k)}
{\displaystyle{\sum_{k=1}^{N^{SIG}}}  \hat{a}_k(\hat s, \hat t,\mu^2, Q^2)
\frac{\Delta f_a^{\gamma^*}(x^k_a,\mu^2)}
{f_a^{\gamma^*}(x^k_a,\mu^2)}}\ ,
\end{equation}
%*************************************************************************
where the sum is over all MC hadrons $k$ in the $p_T$ range 1 GeV $< p_T <$ 2.5 GeV.
This average determines the average $\langle x \rangle$ of the distribution probed 
by this measurement using the mean value theorem for integration, i.e., 
$\langle x \rangle$ is the value of $x$ at which 
$\langle \frac{\Delta g}{g} \rangle \equiv \frac{\Delta g}{g}(\langle x\rangle)$.

%*************************************************************************
\FIGURE{
  \includegraphics*[width=10.0cm]{./figs/deltaGx_final.epsi}
  \caption{\label{fg:DGx_final} The light gray band shows the
   total uncertainty of $\frac{\Delta g}{g}(x)$ vs. $x$ with the 
   statistical and total systematic uncertainty (see Table \protect 
   \ref{tb:syser_anti_d_II}) added in quadrature. Note that the total 
   systematic 
   uncertainty contains a component accounting for the difference between 
   fct.~1 and fct.~2. The point shown represents 
   $\frac{\Delta g}{g}(\langle x\rangle)$ at $\langle x\rangle=0.22$.
   The inner error bar represents the statistical 
   uncertainty and the outer the total uncertainty obtained by adding
   statistical and total systematic uncertainty in quadrature. 
   }
}
%*************************************************************************
Figure \ref{fg:DGx_final} shows the total uncertainty (light gray band) 
of $\frac{\Delta g}{g}(x)$ vs. $x$ evaluated with fct.~1 in the $p_T$ 
range 1.0 GeV $< p_T <$ 2.5 GeV and $\frac{\Delta g}{g}(\langle x \rangle)$.
% and with fct.~1, chosen as 
%default (see Sect.~\ref{sec:SysErr} for systematic uncertainties).
The difference between fct.~1 and fct.~2 is assigned as an additional 
systematic uncertainty on the results from fct.~1 included 
in $sys-models$ (see Sect.~\ref{sec:SysErr}).
The values of $\langle x \rangle$ determined from the two 
functions differ by only 0.007. 

The value for the gluon polarization extracted for the anti-tagged category 
from the deuterium target at $\langle x\rangle = 0.22$ and a scale 
$\langle \mu^2 \rangle$ = 1.35 GeV$^2$ is \\ 
\centerline{ $\mathbf{\frac{\Delta g}{g}(\langle x\rangle, \langle \mu^2\rangle) = 
0.049\pm 0.034 (stat) \pm 0.010 (sys\textrm{-}exp)^{+0.126}_{-0.099}
(sys\textrm{-}models)}$.} \\ 
The scale $\langle \mu^2 \rangle$ was determined by averaging over the scale of all 
signal MC events. The details on the systematic uncertainties are listed in 
table~\ref{tb:syser_anti_d_II}. 

%%%%%%%%%%%%%%%%%%%%%%%%%%%%%%%%%%%%%%%%%%%%%%%%%%%%%%%%%%%%%%%%
%%%%%%%%%%      subsection{Systematic uncertainties}    %%%%%%%%
%%%%%%%%%%%%%%%%%%%%%%%%%%%%%%%%%%%%%%%%%%%%%%%%%%%%%%%%%%%%%%%%
%%%%%%%%%%%%%%%%%%%%%%%%%%%%%%%%%%%%%%%%%%%%%%%%%%%%%%%%%%%%%%%%
\subsection{Systematic uncertainties of $\frac{\Delta g}g$}
\label{sec:SysErr}
%%%%%%%%%%%%%%%%%%%%%%%%%%%%%%%%%%%%%%%%%%%%%%%%%%%%%%%%%%%%%%%%
\subsubsection{\Pythia, \Jetset, and helicity-dependent (averaged) PDFs}
%%%%%%%%%%%%%%%%%%%%%%%%%%%%%%%%%%%%%%%%%%%%%%%%%%%%%%%%%%%%%%%%
At present there is no Monte Carlo code available beyond leading
order that models all subprocesses relevant at the kinematics of this 
experiment. 
Therefore this analysis is limited to leading order. 
As explained in Sect.~\ref{sec:MCtune}, the \Pythia\ model was significantly 
improved to better describe the \hermes\ helicity-averaged data over a wide 
kinematic range. The model contributions to the systematic uncertainty 
($`sys-models'$) are determined by varying the parameters controlling 
the helicity-averaged and helicity-dependent PDFs, the \Pythia\ subprocess cross 
sections and \Jetset\ fragmentation process, and the low-$p_T$ asymmetry. 
An individual uncertainty contribution is determined as the difference between  
\DGG\ with the standard setting and \DGG\ obtained with the alternate setting.
Related types of uncertainties are grouped in classes: `parton distribution 
functions', `\Pythia ~parameters', `low-$p_T$ asymmetry', and 
`fit function fct.~2'.
All the individual and combined uncertainties are shown in 
table~\ref{tb:syser_anti_d_II}.

For most types of uncertainties within a class, \textit{e.g.}, 
helicity-dependent nucleon PDFs, the uncertainty is conservatively estimated 
to be the maximum deviation appearing among the alternative models tested.
Within a class these maximum differences are added in quadrature to form the 
`Total' uncertainty for each class.
The `Total sys-models' uncertainty is obtained by adding those of all 
classes linearly, because of the complexity of correlations between them.

Each of the classes investigated is motivated and discussed below.
\begin{itemize}
\item Parton distribution functions\\
   {\it Spin-dependent nucleon PDFs}\\
   The alternative parameterizations for the quark helicity distributions to GRSV, 
   which were used are: 
   GS-B~\cite{ref:GS-B}, BB-06~\cite{pdf:BB2006}, which includes 
   the most recent $g_1$ data from Refs.~\cite{hermes:g1-06,compass:g1}, 
   and the GRSV standard scenario~\cite{pdf:grsv2000}. 
   The GS-B and BB-06 parameterizations result in
   deviations, which are the second largest systematic uncertainty. They
   are of opposite sign and similar magnitude. The third alternative has a
   negligible effect.\\
   {\it Spin-dependent photon PDFs} \\
   Alternative parameterizations chosen are the  maximum and  minimum
   scenarios of GRS~\cite{ref:GRS-PG,ref:GRS-PG2}.
   The resulting deviations are of opposite sign and similar magnitude and
   make a significant contribution to the overall uncertainty.\\
   {\it Spin-averaged nucleon PDFs} \\
   Using the alternative parameterization GRV98~\cite{pdf:grv98} for the 
   spin-averaged quark and gluon distributions results in a small deviation.\\
   {\it Spin-averaged photon PDFs} \\
   The alternative parameterization is GRS~\cite{pdf:grs1999}, which
   results in a small deviation.
\item \Pythia\ parameters\\
According to the discussion in Sect.~\ref{sec:kt} the following \Pythia\ and 
\Jetset\ parameters have been varied around their central values:
PARP(91) and PARP(99), respectively the initial $k_T$ of the 
partons in the nucleon and photon,  are varied together; 
PARJ(21), which regulates $p_T^{frag}$, where the upper and lower 
values correspond 
to an increase of $\chi_{std}^2$ by 1 unit with respect to the standard 
setting~\cite{Hillenbrand:2005};
PARP(34), which is the multiplicative factor for the factorization 
and renormalization 
scales (PARP(34) $\mu^2$). The scale in the calculation of the 
asymmetries was varied accordingly. 
In order to vary the subprocess fractions directly PARP(90) was varied. 
It is a parameter regulating the cutoff
$p_{T_{min}}={PARP(81)}\left(\frac{W}{PARP(89)}\right)^{PARP(90)}$ between 
direct and anomalous processes as well as soft and hard GVMD processes. 

The combined uncertainty 
of this class is comparable to that from the PDFs class, with PARJ(21) being 
the largest single contribution.
\item Low-$p_T$ asymmetry\\
The logical alternative to $A_{low-p_T}=D\cdot A_1$ (which fits 
the HERMES low-$p_T$ data) is to assume that at low $p_T$ any spin dependence 
is washed out, \textit{e.g.}, $A_{low-p_T}=0$. Any such reduction of 
the asymmetry would only affect the lowest two $p_T$ bins and could 
only increase $\frac{\Delta g}{g}$.
\item Fit function fct.~2\\
For the $x$ dependence, there is an additional class 
%`fit function fct.~2' 
corresponding to functions with the shape of fct.~2.
%due to the different shapes of the two fit functions.
\end{itemize}

%***************************************************************************
\TABLE{
\caption{\label{tb:syser_anti_d_II}
 Average kinematics and results for \dggx\ and the parameter $p_1$ for fct.~1
 with their statistical and systematic uncertainties, from deuteron data for
 anti-tagged events and for the full range in $p_T$. 
 For most types of uncertainties within a class, \textit{e.g.}, 
helicity-dependent
 nucleon PDFs, the uncertainty is conservatively estimated to be the maximum
 deviation appearing among the alternative models tested.
 Within a class (separated box),these maximum differences are added in 
quadrature to form the
 `Total' uncertainty for each class, and referred to as `Total PDFs' and 
 `Total PYTHIA'.
 These components, the `low-$p_T$ asymmetry' and the `fit function fct.~2'
 uncertainties are added linearly to form the `Total sys-models' uncertainty.}
\begin{tabular}{|c|c|r|r|} \hline
 Kinematics  & $\langle x\rangle_{}$ &  0.217 & \\ 
             & $\langle \mu^2 \rangle$ (GeV$^2$) &  1.353 &  \\ \hline
\multicolumn{2}{|c|}{ } &\multicolumn{1}{c|}{ $\Delta g/g$} &\multicolumn{1}{c|}{ $p_1$}\\ 
\hline 
$x$                & value                  & 0.049 &  -1.283 \\
dependence         & statistical uncertainty& 0.034 &   0.083  \\ 
\hline
\multicolumn{4}{|c|}{Systematic Uncertainties }\\ \hline
Category & Model & $\delta(\Delta g/g)$ & $\delta p1$\\ \hline\hline
helicity-dependent & BB-06    & 0.029 &  0.250\\
Nucleon PDF & GS-B        & 0.007 &  0.055\\
            & GRSV-val    & 0.006 &  0.055\\ \hline
helicity-dependent & GRV(max) & 0.024 &  0.245\\
Photon PDF & GRV(min)     &-0.019 & -0.195\\ \hline
 Nucleon PDF & GRV98      &-0.005 & -0.150\\ \hline
 Photon  PDF & GRS        & 0.004 &  0.055\\ 
\hline \hline
{\sc Pythia} & PARP(90)=0.14  &-0.017 & -0.140\\
parameters & PARP(90)=0.18    & 0.007 &  0.040\\
           & PARP(91/99)=0.36 & 0.002 &  0.010\\
           & PARP(91/99)=0.44 &-0.004 & -0.025\\
           & PARJ(21)=0.38    & 0.021 &  0.170\\
           & PARJ(21)=0.42    &-0.035 & -0.290\\
           & PARP(34)=0.5     &-0.014 & -0.170\\
           & PARP(34)=2.0     & 0.016 &  0.170\\ \hline \hline
 \multicolumn{2}{|c|}{low-$p_T$ asymmetry}& 0.046 & 0.395 \\
 \multicolumn{2}{|c|}{Total PDFs $\pm$}        & 0.038 & 0.385\\
 \multicolumn{2}{|c|}{Total {\sc Pythia} $\pm$}& 0.042 & 0.365\\ 
 \multicolumn{2}{|c|}{fit function fct.~2}     &-0.018 & \\ \hline
 \multicolumn{2}{|c|}{Total sys-models $+$} &0.126 & 1.145\\
 \multicolumn{2}{|c|}{Total sys-models $-$} &0.099 & 0.749\\ \hline
 \multicolumn{2}{|c|}{Experimental Systematic} &0.010 & 0.040\\ \hline
\end{tabular}
}
%************************************************************************

%%%%%%%%%%%%%%%%%%%%%%%%%%%%%%%%%%%%%%%%%%%%%%%%%%%%%%%%%%%%%%%%%%%%%%%%
\subsubsection{Experimental systematics} 
%%%%%%%%%%%%%%%%%%%%%%%%%%%%%%%%%%%%%%%%%%%%%%%%%%%%%%%%%%%%%%%%%%%%%%%%
The experimental systematic uncertainty is dominated by the fractional 
uncertainties in beam and target polarization, as shown in
table~\ref{tb:tar}. They are added in quadrature and amount to 3.9\% for 
the asymmetry and 20\% for $\frac{\Delta g}{g}(\langle x\rangle)$ 
%on the average \DGG\ for 
from the deuterium target 
(shown in tables~\ref{tb:syser_anti_d_II} and~\ref{tb:syser_anti_d_I}). 

Due to the rapid reversal of the target spin orientation ($\approx 90$~s) 
the asymmetry extraction is independent of detector efficiency fluctuations.
Possible false asymmetries due to the luminosity normalization are found 
to be negligible. 

%%%%%%%%%%%%%%%%%%%%%%%%%%%%%%%%%%%%%%%%%%%%%%%%%%%%%%%%%%%%%%%%%%%%%%%%%%%
\subsection{Comparison to world data and models}
%%%%%%%%%%%%%%%%%%%%%%%%%%%%%%%%%%%%%%%%%%%%%%%%%%%%%%%%%%%%%%%%%%%%%%%%%%%
Only a few results obtained in leptoproduction exist on \DGG\ at present
\cite{compass:dg2008,smc:glue,compass:dg2006,compass:DIS08}. 
They were obtained from experiments with widely different kinematics 
and they have different scales $\mu^2$. Therefore, they cannot be easily 
compared. 
Nevertheless, for comparison the measurements are shown
together at their respective $\langle x \rangle$ value,
neglecting the $Q^2$ dependence of \DGG . 
The experimental results shown in Fig.~\ref{fg:dg_all} are all obtained 
in leptoproduction, in LO analyses, although for different final states. 
The \hermes\ result is plotted %in Fig.~\ref{fg:dg_all} 
with a horizontal bar 
indicating the half width at half maximum of the $x$ distribution from 
Fig.~\ref{fg:pT_xG}. Fit function fct.~1 is shown for the full $x$ range 
spanned by the \hermes\ data (see Figs.~\ref{fg:dg_all} 
and \ref{fg:MII_funcs}). 
%Fit function fct.~1 is also shown in this $x$ range.
The statistical precision of the \hermes\ result is the best currently
available. The published {\sc Compass} result for 
high-$p_T$ hadron pairs in the region $Q^2<1$ GeV$^2$~\cite{compass:dg2006}
has almost twice the statistical uncertainty.
Concerning the systematic uncertainty the \hermes\ result is solidly 
based on varying many parameters of the well-tuned \Pythia\ simulation and a 
comparison of results from several event categories and targets. The other 
results on \DGG~\cite{compass:dg2008,smc:glue,compass:DIS08} are characterized 
by much larger statistical uncertainties. The earlier HERMES result of \protect \cite{hermes:old-glue} is omitted, because the model used in this paper neglects important underlying subprocesses contributing to the signal and background asymmetries in the kinematic region used to extract \dggx\, and no systematic uncertainty for the model used was determined. Altogether, the presently available 
experimental information from leptoproduction clearly indicates small values 
of \DGG\ over the covered $x$ range.
This conclusion is consistent with the most recent results from
polarized $pp$ collisions from \phenix\ \cite{phenix:Adl} and \star\ 
\cite{star2006}. 
%
%***************************************************************************
\FIGURE{
 \includegraphics[width=14.0cm]{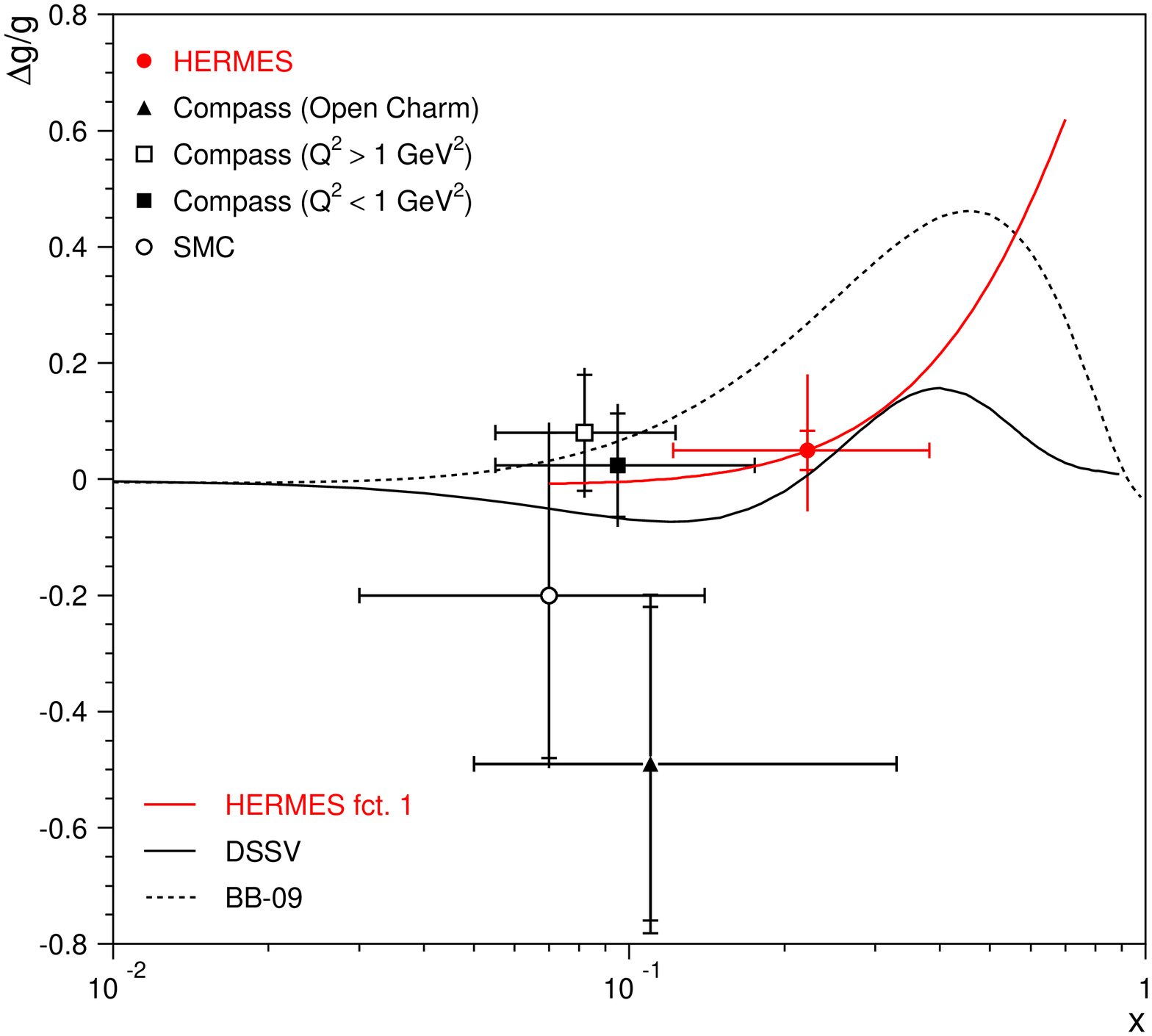}
 \caption{\label{fg:dg_all} The gluon polarization 
  $\frac{\Delta g}{g}(\langle x\rangle)$ from \hermes\ extracted with
  fct.~1 ($\langle x\rangle = 0.22, \langle \mu^2\rangle = 1.35$\,GeV$^2$) 
  compared to the ones from 
  \compass~\cite{compass:dg2006,compass:DIS08,compass:dg2008} 
  (low Q$^2$: $\mu^2=3$ GeV$^2$, high Q$^2$: $\mu^2=2.4$ GeV$^2$, 
  open charm: $\mu^2=13$ GeV$^2$) and SMC~\cite{smc:glue} ($\mu^2=3.6$ GeV$^2$)
  including statistical uncertainties (inner error bars) and total 
  uncertainties (outer error bars). The $x$ region of the data is indicated by
  the horizontal bars. 
  Fit function fct.~1 
  is shown over the full $x$ range spanned by the \hermes\ data. 
  Also shown are a sample of curves from NLO pQCD fits 
  DSSV, and BB-09) at $\mu^2=1.5$ GeV$^2$. For clarity only the central 
  values are shown.}
}
%***************************************************************************

Also shown in Fig.~\ref{fg:dg_all} is \dgg\ calculated from two NLO pQCD fits,
obtained as the ratio of the helicity-dependent PDFs 
(DSSV~\cite{th:DSSV2008}, and BB-09~\cite{pdf:BB2009} ) to the
helicity-averaged PDFs (MRST02~\cite{pdf:mrst02} and DR-09~\cite{pdf:dr09}, 
respectively). The BB-09 NLO pQCD fit is based on the inclusive DIS world 
data set. The DSSV NLO pQCD fit includes the world data on inclusive,  
semi-inclusive DIS and polarised proton proton scattering.
The existing data obtained in leptoproduction on \dggx\ disfavor large 
magnitudes of the gluon polarization over the measured $x$ range, 
in agreement with NLO-QCD fits by DSSV~\cite{th:DSSV2008}. 

%%%%%%%%%%%%%%%%%%%%%%%%%%%%%%%%%%%%%%%%%%%%%%%%%%%%%%%%%%%%%%%%%%%%%%%%%%%
\section{Summary}
\label{sec:conclusion}
%%%%%%%%%%%%%%%%%%%%%%%%%%%%%%%%%%%%%%%%%%%%%%%%%%%%%%%%%%%%%%%%%%%%%%%%%%%
The gluon polarization in the nucleon has been determined 
%at $\la x \ra = 0.22$
by measuring the longitudinal double-spin asymmetry of high-$p_T$
electro-produced single inclusive hadrons at \hermes\ on a deuterium target. 
The value of \DGG\ has been extracted using the measured asymmetries 
along with the subprocesses fractions, asymmetries and kinematics for the 
signal and background processes calculated using the leading-order 
\Pythia\ Monte Carlo code. The value of \DGG\ is obtained from the 
product of gluon polarization and subprocess asymmetries summed over a wide 
range in $x$. The values of \dggp\ and \dggx\ were determined independently.
The systematic uncertainty was evaluated by varying \Pythia\ parameters and 
models of background asymmetries. The final result for the average gluon
polarization in the $p_T$ range 1.0 GeV $< p_T <$ 2.5 GeV is
$\frac{\Delta g}{g}(\langle x\rangle, \langle \mu^2\rangle) = 0.049\pm 0.034 (stat) \pm 0.010
(sys\textrm{-}exp)^{+0.126}_{-0.099}(sys\textrm{-}models)$
at $\langle x \rangle = 0.22$ and $\la\mu^2\ra=1.35~{\rm GeV}^2$.
% and \dggp\ =
% $0.055 \pm 0.033(stat) \pm 0.010 (sys-exp)^{+ 0.124}_{-0.078} (sys-models)$
%at  $\langle p_T \rangle =1.20$ GeV.

%%%%%%%%%%%%   ACKNOWLEDGMENTS   %%%%%%%%%%%%%%%
\acknowledgments
We thank J.~ Bl{\"u}mlein, M. Stratmann and W. Vogelsang for 
helpful discussions and one of us RBRC for financial support.
We gratefully acknowledge the \desy\ management for its support and the staff
at \desy\ and the collaborating institutions for their significant effort.
This work was supported by the FWO-Flanders and IWT, Belgium;
the Natural Sciences and Engineering Research Council of Canada;
the National Natural Science Foundation of China;
the Alexander von Humboldt Stiftung;
the German Bundesministerium f\"ur Bildung und Forschung (BMBF);
the Deutsche Forschungsgemeinschaft (DFG);
the Italian Istituto Nazionale di Fisica Nucleare (INFN);
the MEXT, JSPS, and G-COE of Japan;
the Dutch Foundation for Fundamenteel Onderzoek der Materie (FOM);
the U.K.~Engineering and Physical Sciences Research Council, 
the Science and Technology Facilities Council,
and the Scottish Universities Physics Alliance;
the U.S.~Department of Energy (DOE) and the National Science Foundation (NSF);
the Russian Academy of Science and the Russian Federal Agency for 
Science and Innovations;
and the Ministry of Economy and the Ministry of Education and Science of 
Armenia.

%%%%%%%%%%%%%%%%%%%%%%%%%%%%%%%%%%%%%%%%%%%%%%%%%%%%%%%%%%%%%%%%
%%%%%%%%%%                   Appendix                   %%%%%%%%
%%%%%%%%%%%%%%%%%%%%%%%%%%%%%%%%%%%%%%%%%%%%%%%%%%%%%%%%%%%%%%%%
%%%%%%%%%%%%%%%%%%%%%%%%%%%%%%%%%%%%%%%%%%%%%%%%%%%%%%%%%%%%%%%%%%%%%%%%%%%%%
\newpage
\appendix
\section{Cross section and asymmetries}
\label{ap:equations}
%%%%%%%%%%%%%%%%%%%%%%%%%%%%%%%%%%%%%%%%%%%%%%%%%%%%%%%%%%%%%%%%%%%%%%%%%%%%%
The leading-order formulas for helicity-dependent and helicity-averaged cross 
sections for longitudinally polarized virtual photons and partons for PGF, 
QCDC, DIS~\cite{th:poldis} and some QCD $2\rightarrow 2$ subprocess are 
shown below. These are integrated over the azimuthal angle between the 
positron scattering plane and the production plane.
The hard subprocess asymmetry ($\hat{a}$) is given by
%***************************************************************************
\begin{eqnarray}
\label{eq:ahat}
\hat{a}(\hat{s},\hat{t},\mu^2) &= &\Delta\hat{\sigma}/(2\hat{\sigma}), \\
\Delta\hat{\sigma} &=& \sigpolrrhat -\sigpolrlhat, \\
\hat{\sigma} &=& (\sigpolrrhat +\sigpolrlhat)/2,
\end{eqnarray}
%***************************************************************************
with ($+$) denoting that both partons have the same helicity and
($-$) the opposite helicity. The charge of the struck quark is given by $e_q$
in units of the elementary charge. 
The common factor $C$ is $\frac{4\pi^2\alpha_{em}}{Q^2(1-x)}$.
%***************************************************************************
\begin{eqnarray}
\label{eq:DEL-DIS}
{\rm \bf DIS:} \nonumber \\
\frac{d^2\hat{\sigma}_L^{eq \rightarrow q}}{d\nu dQ^2} &=&  0,\\
\frac{d^2\hat{\sigma}_T^{eq \rightarrow q}}{d\nu dQ^2} &=&  \Gamma C e_q^2x,\\
\frac{d^2\Delta \hat{\sigma}^{eq \rightarrow q}}{d\nu dQ^2} &=& \Gamma C 2e_q^2x. 
\end{eqnarray} 

%***************************************************************************
For PGF and QCDC, there are helicity-averaged transverse and longitudinal, 
as well as a helicity-dependent transverse hard cross sections, note in the following
$d\hat{\sigma}_i \equiv \frac{d^2\hat{\sigma}_i} {d\hat{s} d\hat{t}}$ 
%***************************************************************************
\begin{eqnarray}
\label{eq:DEL-PGF}
{\rm \bf PGF:} \nonumber \\
d\hat{\sigma}_L^{\gamma^* g \rightarrow q \bar{q}} &=& C 
 \frac{\alpha_s e_q^2 }{ 4\pi}\frac{1}{(Q^2+\hat{s})^2}
 \frac{8Q^2\hat{s}}{(\hat{s}+Q^2)^2}, \\
d\hat{\sigma}_T^{\gamma^* g \rightarrow q \bar{q}} &=& C
\frac{\alpha_s e_q^2}{4\pi}\frac{1}{(Q^2+\hat{s})^2}
\left[\frac{Q^4+\hat{s}^2}{
(\hat{s}+Q^2)^2}\frac{\hat{u}^2 +\hat{t}^2 }{\hat{u}\hat{t}}\right],\\
d\Delta \hat{\sigma}_T^{\gamma^* g \rightarrow q \bar{q}} 
&=&  C\frac{\alpha_s e_q^2}{2\pi} \frac{1}{(Q^2+\hat{s})^2}
\left[\frac{Q^2-\hat{s}}{\hat{s}+Q^2} \,
\frac{\hat{u}^2 +\hat{t}^2 }{\hat{u}\hat{t}}\right],
\end{eqnarray}
%***************************************************************************
where $\hat{s}=(p_q+p_{\bar{q}})^2$,
$\hat{t}=(q-p_q)^2,$ $\hat{u}=(q-p_{\bar{q}})^2$
and  $q, p_q, p_{\bar{q}}$ are the 4-momenta of photon, 
final quark and anti-quark, as shown in Fig.~\ref{fg:qcddiag}b.
%***************************************************************************
\begin{eqnarray}
\label{eq:DEL-QCDC}
{\rm \bf QCDC:} \nonumber \\
d\hat{\sigma}_L^{\gamma^* q \rightarrow q g} 
&=&  C\frac{2\alpha_s e_q^2}{3\pi} \frac{1}{(Q^2+\hat{s})^2}
 \frac{4Q^2\hat{u}}{(\hat{s}+Q^2)^2},\\
d\hat{\sigma}_T^{\gamma^* q \rightarrow q g} 
&=&  C\frac{2\alpha_s e_q^2}{ 3\pi} \frac{1}{(Q^2+\hat{s})^2}
\left[2 - \frac{2\hat{u}Q^2}{(\hat{s}+Q^2)^2}-
 \frac{Q^4+\hat{u}^2}{\hat{s}\hat{t}}\right],\\
d\Delta \hat{\sigma}_T^{\gamma^* q \rightarrow q g} 
&=&  C\frac{4\alpha_s e_q^2 }{ 3\pi} \frac{1}{ (Q^2+\hat{s})^2}
\left[\frac{2(Q^2-\hat{u})}{ Q^2+\hat{s}} -
\frac{Q^4+\hat{u}^2}{ \hat{s}\hat{t}}\right].
\end{eqnarray}
%***************************************************************************
For hard QCD $2\rightarrow2$ processes the formulas~\cite{th:babcock}
relevant in this analysis are given in Tab.~\ref{tb:qcd22}.
%***************************************************************************
\TABLE{
\caption{Subprocess differential cross sections 
 $(\Delta) \hat{\sigma}^{ab\rightarrow cd}$ for parton-parton interactions. 
 The common factor of $\frac{\pi\alpha_s^2}{\sh2}$ has been omitted.}
\label{tb:qcd22}
\begin{tabular}{|l|c|c|}
\hline
Reaction & $d\hat{\sigma}/d\hat{t}$ & $d\Delta\hat{\sigma}/d\hat{t}$ \\
\hline
$qg\rightarrow qg$ & 
   $(\sh2+\uh2)[\frac{1}{\th2} -\frac{4}{9\hat s\hat u}]$ &
      $2(\uh2-\sh2)[\frac{4}{9\hat s\hat u} -\frac{1}{\th2}]$\\

$\bar{q}g\rightarrow \bar{q}g$ & 
   $(\sh2+\uh2)[\frac{1}{\th2} -\frac{4}{9\hat s\hat u}]$ &
%   $\frac{\uh2+\sh2}{\th2} -\frac{4}{9}\frac{\sh2+\uh2}{\hat s\hat u}$ &
      $2(\uh2-\sh2)[\frac{4}{9\hat u\hat s} -\frac{1}{\th2}]$\\

$gg\rightarrow q\bar q$ & 
   $\frac{\uh2+\th2}{6\hat u\hat t} -\frac{3}{8}\frac{\th2+\uh2}{\sh2}$ &
       $ \frac{3}{4}\frac{\th2+\uh2}{\sh2} -\frac{\uh2+\th2}{3\hat u\hat t}$ \\

$gg\rightarrow gg$ & 
  $\frac{9}{2}(3-\frac{\hat t\hat u}{\sh2}-\frac{\hat s\hat u}{\th2}-\frac{\hat s\hat t}{\uh2})$ &
  $9(-3 +2\frac{\sh2}{\hat u\hat t} + \frac{\hat u\hat t}{\sh2})$\\
\hline
$q_aq_b\rightarrow q_aq_b$ &
 $\frac{4}{9}[\frac{\sh2 +\uh2}{\th2} + \delta_{ab}(\frac{\sh2 +\th2}{\uh2}- \frac{2\sh2}{3\hat t\hat u})]$&
  $\frac{8}{9}[\frac{\sh2-\uh2}{\th2} - \delta_{ab}(\frac{\th2-\sh2}{\uh2}+\frac{2\sh2}{3\hat t\hat u})]$ \\

$q_a\bar{q}_b\rightarrow q_c\bar{q}_d$ &
  $\frac{4}{9}[\delta_{ac}\delta_{bd}\frac{\uh2}{\th2} +\delta_{cd}\delta_{ab}\frac{\th2+\uh2}{\sh2} -$ &
   $\frac{8}{9}[-\delta_{ac}\delta_{bd}\frac{\uh2}{\th2} -\delta_{cd}\delta_{ab}\frac{\th2+\uh2}{\sh2}+ $ \\
 & $\delta_{ad}\delta_{cd}\frac{2\uh2}{3\hat{s}\hat{t}}+\delta_{ab}\delta_{bd}\frac{\sh2}{\th2}]$ &
    $\delta_{ad}\delta_{cd}\frac{2\uh2}{3\hat{s}\hat{t}}+\delta_{ab}\delta_{bd}\frac{\sh2}{\th2}]$ \\

$q\bar{q}\rightarrow gg$ &
   $\frac{32}{27}\frac{\th2+\uh2}{\hat{u}\hat{t}} -\frac{8}{3}\frac{\th2+\uh2}{\sh2}$ &
     $-\frac{64}{27}\frac{\th2+\uh2}{\hat{u}\hat{t}} +\frac{16}{3}\frac{\th2+\uh2}{\sh2}$   \\
\hline
\end{tabular}
}
%***************************************************************************

For lepton scattering, the helicity-dependent and helicity-averaged cross sections 
are given by
%***************************************************************************
\begin{eqnarray}
\frac{d^4 \Delta \sigma^{eq\rightarrow e'ff}}{d\nu dQ^2 d\hat{s} d\hat{t}} &=& 
D\Gamma_T \frac{d^2 \Delta \hat{\sigma}_T^{\gamma^* q \rightarrow f f}}{d\hat{s} d\hat{t}}, \\
\frac{d^4\sigma^{eq\rightarrow e'ff}}{d\nu dQ^2 d\hat{s} d\hat{t}} &=& 
\Gamma_T ( \frac{d^2\hat{\sigma}_T^{\gamma^* q \rightarrow f f}}{d\hat{s} d\hat{t}} +
 \epsilon \cdot \frac{d^2\hat{\sigma}_L^{\gamma^* q \rightarrow f f}}{d\hat{s} d\hat{t}} ),  
\end{eqnarray}
%***************************************************************************
where $f$ stands for a quark, an antiquark or a gluon in the final state and 
$D$ is the virtual-photon depolarization factor
%***************************************************************************
\begin{equation}
\label{depolfact}
 D(y,Q^2) = \frac{y\left[\left(1+\gamma^2y/2\right)(2-y)-2y^2m_e^2/Q^2\right]}
{y^2\left(1-2m_e^2/Q^2\right)(1+\gamma^2)+2(1+R)
\left(1-y-\gamma^2y^2/4\right)},
\end{equation}
%***************************************************************************
$\gamma^2=Q^2/\nu^2$, $R=\sigma_L/\sigma_T$ for the subprocess, and 
$\Gamma_T$ is the transverse photon flux factor
%***************************************************************************
\begin{eqnarray}
\label{photonflux}
\Gamma_T&=& \frac{\alpha_{em}(1-x)}{2\pi Q^2\nu}\left[y^2
\left(1-2m_e^2/Q^2\right)
+\frac {2\left(1-y-\gamma^2y^2/4\right)}{1+\gamma^2}\right],
\end{eqnarray}
%***************************************************************************
and 
%***************************************************************************
\begin{equation} 
\epsilon = \left[1+\frac12\left(1-2m_e^2/Q^2\right)
\frac{y^2+\gamma^2y^2}{1-y-\gamma^2y^2/4}\right]^{-1}.
\label{epsilonflux}
\end{equation}
%***************************************************************************

%%%%%%%%%%%%%%%%%%%%%%%%%%%%%%%%%%%%%%%%%%%%%%%%%%%%%%%%%%%%%%%%%%%%%%%%%%%%%%
%%%%%%%%%%%%%%%%%%%%%%%%%%%%%%%%%%%%%%%%%%%%%%%%%%%%%%%%%%%%%%%%%%%%%%%%%%%%%%
\newpage
\section{Tuned \Pythia\ parameters}
\label{sec:pythia_param}
%%%%%%%%%%%%%%%%%%%%%%%%%%%%%%%%%%%%%%%%%%%%%%%%%%%%%%%%%%%%%%%%%%%%%%%%%%%%%%
\TABLE[h]{
\caption{The PYTHIA parameters, tuned to HERMES data, which are different
 from the default settings that can be found in Ref.~\cite{PYTHIA6.2}.}
 \label{tb:pystd}
\begin{tabular}{|llll|}
\hline
MSEL=2      &
MSTP(13)=2  &
MSTP(17)=6  &
MSTP(20)=4  \\
MSTP(38)=4  &
MSTP(61)=0  &
MSTP(71)=0  &
MSTP(81)=0  \\
MSTP(92)=4  &
MSTP(101)=1 &
MSTP(121)=1 & \\ \hline
PARP(2)=7 &
PARP(18)=0.17  &
PARP(91)=0.40  &
PARP(93)=2     \\
PARP(99)=0.40  &
PARP(102)=0.5  &
PARP(103)=0.5  &
PARP(104)=0.3  \\
PARP(111)=0    &
PARP(121)=2    &
PARP(161)=3.00 &
PARP(162)=24.6 \\
PARP(163)=18.8  &
PARP(165)=0.477 &
PARP(166)=0.676 & \\ \hline
PARJ(1)=0.029  &
PARJ(2)=0.283  &
PARJ(3)=1.20   &
PARJ(21)=0.40  \\
PARJ(41)=1.94  &
PARJ(42)=0.544 &
PARJ(45)=1.05  & \\ \hline
MSTJ(12)=1     &
MSTJ(45)=4     &
               & \\
MSTU(112)=4    &
MSTU(113)=4    &
MSTU(114)=4    & \\ \hline
CKIN(1)=1.0    &
CKIN(65)=$1.\cdot 10^{-9}$ &
CKIN(66)=100.  & \\ \hline
\end{tabular}
}
%%%%%%%%%%%%%%%%%%%%%%%%%%%%%%%%%%%%%%%%%%%%%%%%%%%%%%%%%%%%%%%%%%%%%%%%%%%%%%

\newpage

%%%%%%%%%%%%%%%%%%%%%%%%%%%%%%%%%%%%%%%%%%%%%%%%%%%%%%%%%%%%%%%%%%%%%%%%%%%%%%
\newpage
\section{Systematic Uncertainties}
\label{systables}
%%%%%%%%%%%%%%%%%%%%%%%%%%%%%%%%%%%%%%%%%%%%%%%%%%%%%%%%%%%%%%%%%%%%%%%%%%%%%%
\TABLE{
\caption{\label{tb:syser_anti_d_I}
  Average kinematics and results for \dggp\ with their statistical and
  systematic uncertainties, from deuteron data for anti-tagged events shown
  for the four bins and the full range in $p_T$.
  For most types of uncertainties within a class, \textit{e.g.}, spin-dependent
  nucleon PDFs, the uncertainty is conservatively estimated to be the maximum
  deviation appearing among the alternative models tested.
  Within a class (separated box), these maximum differences are added in 
  quadrature to form the
  `Total' uncertainty for each class, and referred to as `Total PDFs' and
  `Total PYTHIA'.
  These components and the `$A^{low-p_T}$' uncertainties are added linearly
  into the `Total sys-Models' uncertainty.}
\begin{tabular}{|c|c|r|r|r|r||r|} \hline
           & $p_T$ bin (GeV) & 1.0-1.2 & 1.2-1.5 & 1.5-1.8&1.8-2.5& 1.0-2.5\\ \hline
Kinematics & $\la p_T\ra$ (GeV) & 1.11 & 1.30 & 1.60 & 1.90 & 1.20 \\ \hline

\multicolumn{7}{|c|}{Values}\\ \hline %\tabmspace
$p_T$  & $\Delta g/g$               & 0.017 & 0.033 & 0.026 & 0.619 & 0.055 \\
dependence  & $\delta(\Delta g/g)(stat)$ &  0.067 & 0.046 & 0.073 & 0.154 & 0.033 \\ \hline
\multicolumn{7}{|c|}{Systematic Uncertainties }\\ \hline
Category & Model & \multicolumn{5}{c}{$\delta(\Delta g/g)$}\vline\\ \hline
spin-dependent & BB-06     &$ 0.038$&$ 0.019$&$ 0.026$&$ 0.089$&$ 0.027$ \\
Nucleon PDF & GS-B         &$-0.030$&$-0.031$&$-0.045$&$-0.104$&$-0.037$ \\
            & GRSV-val     &$ 0.007$&$ 0.005$&$ 0.008$&$ 0.001$&$ 0.006$ \\ \hline
spin-dependent & GRV(max)  &$ 0.027$&$ 0.019$&$ 0.014$&$ 0.060$&$ 0.025$ \\
Photon PDF & GRV(min)      &$-0.019$&$-0.016$&$-0.013$&$-0.069$&$-0.020$\\ \hline
 Nucleon PDF & GRV98       &$ 0.020$&$-0.008$&$-0.031$&$-0.144$&$-0.006$ \\ \hline
 Photon  PDF & GRS         &$ 0.029$&$ 0.001$&$-0.018$&$ 0.100$&$ 0.004$ \\
\hline  \hline
{\sc Pythia} & PARP(90)=0.14  &$-0.006$&$-0.017$&$ 0.003$&$-0.076$&$-0.016$ \\
Parameters & PARP(90)=0.18    &$ 0.023$&$-0.006$&$-0.018$&$ 0.014$&$ 0.008$ \\
           & PARP(91/99)=0.36 &$ 0.002$&$-0.002$&$ 0.002$&$-0.058$&$-0.001$\\
           & PARP(91/99)=0.44 &$ 0.004$&$-0.009$&$ 0.002$&$-0.011$&$-0.002$ \\
           & PARJ(21)=0.38    &$ 0.036$&$ 0.018$&$ 0.031$&$ 0.110$&$ 0.021$ \\
           & PARJ(21)=0.42    &$-0.023$&$-0.040$&$-0.032$&$-0.072$&$-0.034$ \\
           & PARP(34)=0.5     &$ 0.017$&$-0.006$&$-0.024$&$-0.187$&$-0.014$ \\
           & PARP(34)=2.0     &$-0.013$&$ 0.012$&$ 0.012$&$ 0.210$&$ 0.012$ \\ \hline \hline
 \multicolumn{2}{|c|}{low-p$_T$ asymmetry}&$0.108$&$0.037$&$0.009$&$0.005$&$0.046$ \\
 \multicolumn{2}{|c|}{Total PDFs $\pm$}        &$0.058$&$0.038$&$0.059$&$0.215$&$0.045$ \\
 \multicolumn{2}{|c|}{Total {\sc Pythia}$\pm$} &$0.047$&$0.038$&$0.047$&$0.258$&$0.033$ \\
\hline
 \multicolumn{2}{|c|}{Total sys-Models $+$}
&$0.212 $&$0.113$&$0.116 $&$0.473 $&$0.124 $ \\
 \multicolumn{2}{|c|}{Total sys-Models $-$}
&$0.105 $&$0.076$&$0.107$&$0.477$&$0.078 $ \\ \hline
 \multicolumn{2}{|c|}{Experimental Systematic} &$0.003 $&$0.006$&$0.009$&$0.002$&$0.011$ \\ \hline
\end{tabular}\\
}
\clearpage
%%%%%%%%%%%%%%%%%%%%%%%%%%%%%%%%%%%%%%%%%%%%%%%%%%%%%%%%%%%%%%%%%%%%%%%%%%%%%%

\newpage

%%%%%%%%%%%%%%%%%%%%%%%%%%%%%%%%%%%%%%%%%%%%%%%%%%%%%%%%%%%%%%%%%%%%%%%%%%%%%%
\newpage
\section{Hadron Asymmetries}
\label{asmtables}
%%%%%%%%%%%%%%%%%%%%%%%%%%%%%%%%%%%%%%%%%%%%%%%%%%%%%%%%%%%%%%%%%%%%%%%%%%%%%%
\TABLE[h]{
\caption{\label{tb:Anti-tagged}
 Anti-tagged inclusive hadrons: measured longitudinal double-spin
 asymmetry for positive and negative hadrons on a deuterium (hydrogen) target.
 The uncertainty shown is statistical only. There is an additional overall 
 normalization uncertainty of 3.9\% (5.2\%), all other
 systematic uncertainties are negligible.}
\begin{tabular}{|c|c|r|r|r|r|} \hline 
 & & \multicolumn{4}{c|}{}\\[-1em]
 $p_T$-bin & $\langle p_T\rangle$ & \multicolumn{2}{c|}{deuterium} & 
 \multicolumn{2}{c|}{hydrogen} \\
(GeV) & (GeV) & $A^{meas}(h^+)$ & $A^{meas}(h^-)$ & 
                $A^{meas}(h^+)$ & $A^{meas}(h^-)$ \\ \hline
0.00--0.15 & 0.12  &  0.0002 $\pm$  0.0011 & -0.0028 $\pm$ 0.0011 
                   &  0.0152 $\pm$  0.0018 &  0.0109 $\pm$ 0.0020 \\
0.15--0.30 & 0.23  & -0.0013 $\pm$  0.0004 & -0.0023 $\pm$ 0.0004 
                   &  0.0107 $\pm$  0.0007 &  0.0081 $\pm$ 0.0008 \\
0.30--0.45 & 0.37  & -0.0032 $\pm$  0.0004 & -0.0042 $\pm$ 0.0004 
                   &  0.0098 $\pm$  0.0007 &  0.0080 $\pm$ 0.0008 \\
0.45--0.60 & 0.52  & -0.0029 $\pm$  0.0006 & -0.0055 $\pm$ 0.0006 
                   &  0.0101 $\pm$  0.0009 &  0.0095 $\pm$ 0.0011 \\
0.60--0.75 & 0.67  & -0.0038 $\pm$  0.0008 & -0.0037 $\pm$ 0.0009 
                   &  0.0127 $\pm$  0.0014 &  0.0119 $\pm$ 0.0017 \\
0.75--0.90 & 0.81  &  0.0005 $\pm$  0.0012 & -0.0027 $\pm$ 0.0015 
                   &  0.0146 $\pm$  0.0021 &  0.0105 $\pm$ 0.0026 \\
0.90--1.05 & 0.96  & -0.0003 $\pm$  0.0019 & -0.0007 $\pm$ 0.0023 
                   &  0.0166 $\pm$  0.0033 &  0.0088 $\pm$ 0.0043 \\
1.05--1.20 & 1.11  &  0.0069 $\pm$  0.0032 & -0.0033 $\pm$ 0.0038 
                   &  0.0351 $\pm$  0.0055 &  0.0091 $\pm$ 0.0071 \\
1.20--1.35 & 1.26  &  0.0150 $\pm$  0.0054 & -0.0021 $\pm$ 0.0063 
                   &  0.0563 $\pm$  0.0094 &  0.0167 $\pm$ 0.0118 \\
1.35--1.50 & 1.41  &  0.0174 $\pm$  0.0091 &  0.0062 $\pm$ 0.0104 
                   &  0.0487 $\pm$  0.0157 &  0.0035 $\pm$ 0.0197 \\
1.50--1.65 & 1.56  &  0.0429 $\pm$  0.0148 & -0.0017 $\pm$ 0.0172 
                   &  0.0886 $\pm$  0.0256 & -0.0759 $\pm$ 0.0327 \\
1.65--1.80 & 1.71  &  0.0719 $\pm$  0.0238 & -0.0001 $\pm$ 0.0277 
                   &  0.1317 $\pm$  0.0412 & -0.0398 $\pm$ 0.0530 \\
1.80--2.00 & 1.88  & -0.0075 $\pm$  0.0342 & -0.0027 $\pm$ 0.0410 
                   &  0.1605 $\pm$  0.0596 &  0.0428 $\pm$ 0.0776 \\
2.00--2.50 & 2.16  &  0.0377 $\pm$  0.0463 & -0.0908 $\pm$ 0.0572 
                   &  0.1071 $\pm$  0.0807 &  0.0575 $\pm$ 0.1128 \\
2.50--5.00 & 3.04  & -0.0071 $\pm$  0.0817 &  0.1201 $\pm$ 0.0960 
                   &  0.0112 $\pm$  0.1405 & -0.0102 $\pm$ 0.1546 \\ \hline 
\end{tabular}
}
%%%%%%%%%%%%%%%%%%%%%%%%%%%%%%%%%%%%%%%%%%%%%%%%%%%%%%%%%%%%%%%%%%%%%%%%%%%%%%
\TABLE[h]{
\caption{\label{tb:Tagged}
 Tagged inclusive hadrons: measured longitudinal double-spin asymmetry 
 for positive and negative  hadrons on a deuterium (hydrogen) target.
 The uncertainty shown is statistical only. There is an additional overall
 normalization uncertainty of 3.9\% (5.2\%), all other
 systematic uncertainties are negligible.}
\begin{tabular}{|c|c|r|r|r|r|} \hline 
 & & \multicolumn{4}{c|}{}\\[-1em]
$p_T$-bin & $\langle p_T\rangle$ & \multicolumn{2}{c|}{deuterium}
& \multicolumn{2}{c|}{hydrogen}\\
(GeV) & (GeV) & $A^{meas}(h^+)$ & $A^{meas}(h^-)$ & 
                $A^{meas}(h^+)$ & $A^{meas}(h^-)$ \\ \hline
0.00--0.15 & 0.10 &  0.0229 $\pm$  0.0045 &  0.0159 $\pm$ 0.0050 
                  &  0.0896 $\pm$  0.0073 &  0.0643 $\pm$ 0.0089 \\
0.15--0.30 & 0.23 &  0.0201 $\pm$  0.0029 &  0.0120 $\pm$ 0.0033 
                  &  0.0850 $\pm$  0.0048 &  0.0605 $\pm$ 0.0058 \\
0.30--0.45 & 0.37 &  0.0188 $\pm$  0.0031 &  0.0134 $\pm$ 0.0036 
                  &  0.0807 $\pm$  0.0053 &  0.0479 $\pm$ 0.0064 \\
0.45--0.60 & 0.52 &  0.0186 $\pm$  0.0040 &  0.0134 $\pm$ 0.0048 
                  &  0.0789 $\pm$  0.0068 &  0.0639 $\pm$ 0.0085 \\
0.60--0.75 & 0.67 &  0.0211 $\pm$  0.0057 &  0.0066 $\pm$ 0.0068 
                  &  0.0661 $\pm$  0.0095 &  0.0494 $\pm$ 0.0122 \\
0.75--1.00 & 0.85 &  0.0104 $\pm$  0.0072 &  0.0106 $\pm$ 0.0088 
                  &  0.0722 $\pm$  0.0122 &  0.0683 $\pm$ 0.0159 \\
1.00--1.30 & 1.11 &  0.0465 $\pm$  0.0143 &  0.0367 $\pm$ 0.0176 
                  &  0.1055 $\pm$  0.0237 &  0.1318 $\pm$ 0.0314 \\
1.30--1.60 & 1.40 &  0.0660 $\pm$  0.0364 & -0.0586 $\pm$ 0.0431 
                  &  0.1410 $\pm$  0.0607 &  0.0481 $\pm$ 0.0813 \\
1.60--2.00 & 1.72 &  0.0165 $\pm$  0.0903 & -0.0390 $\pm$ 0.1086 
                  &  0.0627 $\pm$  0.1501 & -0.0229 $\pm$ 0.2018 \\
2.00--3.50 & 2.18 &  0.1534 $\pm$  0.3059 &  0.3929 $\pm$ 0.3798 
                  & -1.5868 $\pm$  0.9548 &  1.4778 $\pm$ 1.4182 \\ \hline 
\end{tabular}
}
\clearpage
%%%%%%%%%%%%%%%%%%%%%%%%%%%%%%%%%%%%%%%%%%%%%%%%%%%%%%%%%%%%%%%%%%%%%%%%%%%%%%

\newpage

%%%%%%%%%%%%%%%%%%%%%%%%%%%%%%%%%%%%%%%%%%%%%%%%%%%%%%%%%%%%%%%%%%%%%%%%%%%%%%
\newpage
\TABLE[ht]{
\caption{Inclusive hadron pairs: measured longitudinal double-spin
 asymmetry for proton and deuterium targets. The uncertainty shown  
 is statistical only. There is an additional overall normalization  
 uncertainty of 5.2\% (3.9\%) for hydrogen (deuterium), all other
 systematic uncertainties are negligible.}
\label{tb:2hasym}
\hspace*{\textwidth}\\
\begin{tabular}{|c|r|r|} \hline  
& \multicolumn{2}{c|}{}\\[-1em]
$(\sum{p_{T(beam)}^2})_{min}$ & \multicolumn{2}{c|}{$A^{meas}$(hh)} \\
(GeV$^2$) & hydrogen & deuterium\\ \hline
 1.00  &  0.023 $\pm$  0.004 & -0.005 $\pm$ 0.003 \\
 1.20  &  0.018 $\pm$  0.006 & -0.005 $\pm$ 0.003 \\
 1.40  &  0.016 $\pm$  0.008 & -0.000 $\pm$ 0.004 \\
 1.60  &  0.031 $\pm$  0.010 &  0.001 $\pm$ 0.006 \\
 1.80  &  0.047 $\pm$  0.013 & -0.006 $\pm$ 0.007 \\ 
 2.00  &  0.041 $\pm$  0.016 & -0.010 $\pm$ 0.009 \\
 2.50  &  0.084 $\pm$  0.028 & -0.028 $\pm$ 0.015 \\
 3.00  & -0.001 $\pm$  0.045 & -0.041 $\pm$ 0.025 \\
 4.00  &  0.042 $\pm$  0.108 & -0.080 $\pm$ 0.057 \\ \hline
\end{tabular}
}
\clearpage
%%%%%%%%%%%%%%%%%%%%%%%%%%%%%%%%%%%%%%%%%%%%%%%%%%%%%%%%%%%%%%%%%%%%%%%%%%%%%%

%%%%%%%%%%%%%%%%%%%%%%%%%%%%%%%%%%%%%%%%%%%%%%%%%%%%%%%%%%%%%%%%%%%%%%%%%%%%%%
\newpage
\bibliography{paper_v21}
%%%%%%%%%%%%%%%%%%%%%%%%%%%%%%%%%%%%%%%%%%%%%%%%%%%%%%%%%%%%%%%%%%%%%%%%%%%%%%

%%%%%%%%%%%%%%%%%%%%%%%%%%%%%%%%%%%%%%%%%%%%%%%%%%%%%%%%%%%%%%%%%%%%%%%%%%%%%%
\end{document}